\documentclass[pra,aps,twocolumn, nobalancelastpage, superscriptaddress]{revtex4-2}
\usepackage{graphicx}
\usepackage{dcolumn}
\usepackage{bm}
\usepackage{hyperref}
\usepackage{siunitx}
\usepackage{appendix}
\usepackage{amsmath}
\usepackage{amssymb}

\newcommand{\pmulti}{p_{\text{m}}}

\begin{document}

\title[Robust excitation of C-band quantum dots for quantum communication]{Robust excitation of C-band quantum dots for quantum communication}

\author{Michal Vyvlecka}
 \altaffiliation[These authors contributed equally: Michal Vyvlecka, Lennart Jehle. Authors to whom correspondence should be addressed: michal.vyvlecka@univie.ac.at, lennart.jehle@univie.ac.at.]{}
 \affiliation{University of Vienna, Faculty of Physics \& Vienna Doctoral School in Physics \& Vienna Center for Quantum Science and Technology, Boltzmanngasse 5, A-1090 Vienna, Austria}

\author{Lennart Jehle}
\altaffiliation[These authors contributed equally: Michal Vyvlecka, Lennart Jehle. Authors to whom correspondence should be addressed: michal.vyvlecka@univie.ac.at, lennart.jehle@univie.ac.at.]{}
 \affiliation{University of Vienna, Faculty of Physics \& Vienna Doctoral School in Physics \& Vienna Center for Quantum Science and Technology, Boltzmanngasse 5, A-1090 Vienna, Austria}

\author{Cornelius Nawrath}
 \affiliation{Institut f\"ur Halbleiteroptik und Funktionelle Grenzfl\"achen, Center for Integrated Quantum Science and Technology (IQ\textsuperscript{ST}) and SCoPE, University of Stuttgart, Allmandring 3, 70569 Stuttgart, Germany}%

\author{Francesco Giorgino}
\affiliation{University of Vienna, Faculty of Physics \& Vienna Doctoral School in Physics \& Vienna Center for Quantum Science and Technology, Boltzmanngasse 5, A-1090 Vienna, Austria}

\author{Mathieu Bozzio}
\affiliation{University of Vienna, Vienna Center for Quantum Science and Technology, Faculty of Physics, Boltzmanngasse 5, A-1090 Vienna, Austria}

\author{Robert Sittig}
\affiliation{Institut f\"ur Halbleiteroptik und Funktionelle Grenzfl\"achen, Center for Integrated Quantum Science and Technology (IQ\textsuperscript{ST}) and SCoPE, University of Stuttgart, Allmandring 3, 70569 Stuttgart, Germany}

\author{Michael Jetter}
\affiliation{Institut f\"ur Halbleiteroptik und Funktionelle Grenzfl\"achen, Center for Integrated Quantum Science and Technology (IQ\textsuperscript{ST}) and SCoPE, University of Stuttgart, Allmandring 3, 70569 Stuttgart, Germany}

\author{Simone L. Portalupi}
\affiliation{Institut f\"ur Halbleiteroptik und Funktionelle Grenzfl\"achen, Center for Integrated Quantum Science and Technology (IQ\textsuperscript{ST}) and SCoPE, University of Stuttgart, Allmandring 3, 70569 Stuttgart, Germany}

\author{Peter Michler}
\affiliation{Institut f\"ur Halbleiteroptik und Funktionelle Grenzfl\"achen, Center for Integrated Quantum Science and Technology (IQ\textsuperscript{ST}) and SCoPE, University of Stuttgart, Allmandring 3, 70569 Stuttgart, Germany}

\author{Philip Walther}
\affiliation{University of Vienna, Vienna Center for Quantum Science and Technology, Faculty of Physics, Boltzmanngasse 5, A-1090 Vienna, Austria}
\affiliation{Christian Doppler Laboratory for Photonic Quantum Computer, Faculty of Physics, University of Vienna, Vienna, Austria}

\date{\today}

\begin{abstract}
Building a quantum internet requires efficient and reliable quantum hardware, from photonic sources to quantum repeaters and detectors, ideally operating at telecommunication wavelengths.
Thanks to their high brightness and single-photon purity, quantum dot (QD) sources hold the promise to achieve high communication rates for quantum-secured network applications.
Furthermore, it was recently shown that excitation schemes such as longitudinal acoustic phonon-assisted (LA) pumping provide security benefits by scrambling the coherence between the emitted photon-number states.
In this work, we investigate further advantages of LA-pumped quantum dots with emission in the telecom C-band as a core hardware component of the quantum internet.
We experimentally demonstrate how varying the pump power and spectral detuning with respect to the excitonic transition can improve quantum-secured communication rates and provide stable emission statistics regardless of network-environment fluctuations.
These findings have significant implications for general implementations of QD single-photon sources in practical quantum communication networks.
\end{abstract}

\maketitle

The emergence of practical quantum technology paves the way to a quantum internet -- a network of connected quantum computers capable of reaching computational speed-ups in various tasks such as prime factoring~\cite{MLL:NatPhot12}, machine learning~\cite{Saggio2021} and the verification of NP-complete problems with limited information~\cite{Centrone2021}. Although such schemes are appealing, most are technologically challenging, while the security advantages provided by quantum cryptography are more tangible~\cite{Pan:RevMod20,BAL:npjqi17,BBR:PRL18}. A broad range of quantum-cryptographic primitives including quantum key distribution (QKD)~\cite{Pan:RevMod20,BAL:npjqi17,BBR:PRL18}, quantum coin flipping~\cite{NYC:NatComm23,BBB:NC11,PJ+:natcomm14}, unforgeable quantum tokens~\cite{SKS:NatComm23,Kent:npjQI22,GAA:pra18}, and quantum bit commitment ~\cite{NJC:NatComms12,Zbind:PRL13,Pan:PRL14} have been developed to demonstrate some security advantage over their classical counterparts. The success of a future quantum internet then relies on the development of fundamental quantum hardware (sources, repeaters and detectors) which should adhere to these primitives' security standards, provide high communication rates, and operate reliably in a real-world environment~\cite{Kimble:2008}.

Non-classical light sources such as spontaneous parametric down-conversion~\cite{Ren:2017,Ma:2012}, nitrogen-vacancy centers~\cite{Hensen:2015} and trapped atoms~\cite{Ritter:2012}, have been used as hardware for the first quantum networks.
In recent years, semiconductor quantum dots (QDs) have materialized as highly versatile and quality single-photon sources~\cite{Wang2019a,BassoBasset2019,Zopf2019,Tomm:2021aa,Zhai2022}, with outstanding end-to-end efficiencies overcoming $57\,\%$ and the potential to reach repetition rates of tens of GHz~\cite{Tomm:2021aa}.
Such emission properties of QDs have led to the implementation of complex network building blocks relying on quantum teleportation~\cite{AMH:npjQI20,BSS:npjQI21} and quantum entanglement swapping~\cite{BassoBasset2019,Zopf2019,Lodahl:2017}.
Regarding the emission wavelength, the spectral regime of the telecom C-band (\SI{1530}{\nano\meter} to \SI{1565}{\nano\meter}) is highly appealing, due to its global absorption minimum in standard silica fibers, the possibility to implement daylight satellite communication~\cite{Liao2017} and the compatibility with the mature silicon photonic platforms~\cite{Wang2020}.
QDs with emission wavelengths in and around the C-band are available on indium phosphide (InP)~\cite{Miyazawa2016,Anderson2020b,Shooter2020} and gallium arsenide (GaAs) material system~\cite{Lettner2021,Sittig2022,Nawrath2023}, and circumvent the technical overhead and losses of quantum frequency conversion~\cite{vLeent2020}.
Embedded  in circular Bragg cavities, QDs based on the well-established GaAs platform have simultaneously demonstrated high brightness and high purity values recently~\cite{Nawrath2023}.

Previous works have investigated the advantages and drawbacks of various optical pumping schemes (resonant, phonon-assisted and two-photon excitation) in terms of efficiency, single-photon purity and indistinguishability~\cite{PhysRevLett.126.233601, Armando:ACS17, ReindlPRB2019}.
On the other hand, it was recently shown that such schemes must be carefully tuned to satisfy the security assumptions of each quantum-cryptographic application~\cite{BVC:npjQI22}.
Crucially, quantum coherences between the emitted photon-number components must be scrambled for optimal performance, which is inherently provided by some optical pumping schemes such as longitudinal phonon-assisted (LA) excitation and two-photon excitation (TPE)~\cite{BVC:npjQI22}.
On top of their intrinsic security benefits, LA schemes are fairly insensitive to pump instabilities like power or polarization fluctuations, making them suitable for real-life communication networks~\cite{PhysRevLett.126.233601, Armando:ACS17}.
These excitation schemes are also beneficial for QDs with a complex charge environment, while other pumping schemes such as TPE can typically only address charge-neutral transitions.
Moreover, unlike for neutral transitions, charged excitons can enhance polarized emission in polarized cavities, an important feature for most applications\cite{Tomm:2021aa}.
Finally, LA schemes do not require challenging single-photon polarization filtering (contrary to the resonant counterpart), and thus promise an experimentally straightforward way to obtain simultaneously high brightness and 
purity with high reproducibility in quantum dot fabrication and experimental setups~\cite{PhysRevLett.126.233601}.

In this work, we combine all aforementioned advantages of LA excitation in the C-band, and exploit its tunable parameters to investigate the complex dependence of brightness and purity on pump power and spectral detuning.
We illustrate how this non-trivial behavior affects the security of quantum-cryptographic primitives with the example of single-photon QKD, and how the optimal operation conditions depend on the communication distance.
In agreement with theoretical findings~\cite{Axt:PRL19}, our results show that the characteristics of LA excitation can be tuned to achieve the ideal photon-number statistics.
This optimization is reminiscent of the mean-photon adjustment required in weak coherent state (WCS) implementations~\cite{Ma:Thesis08}.

\begin{figure}
	\begin{center}
		\includegraphics[width=85mm]{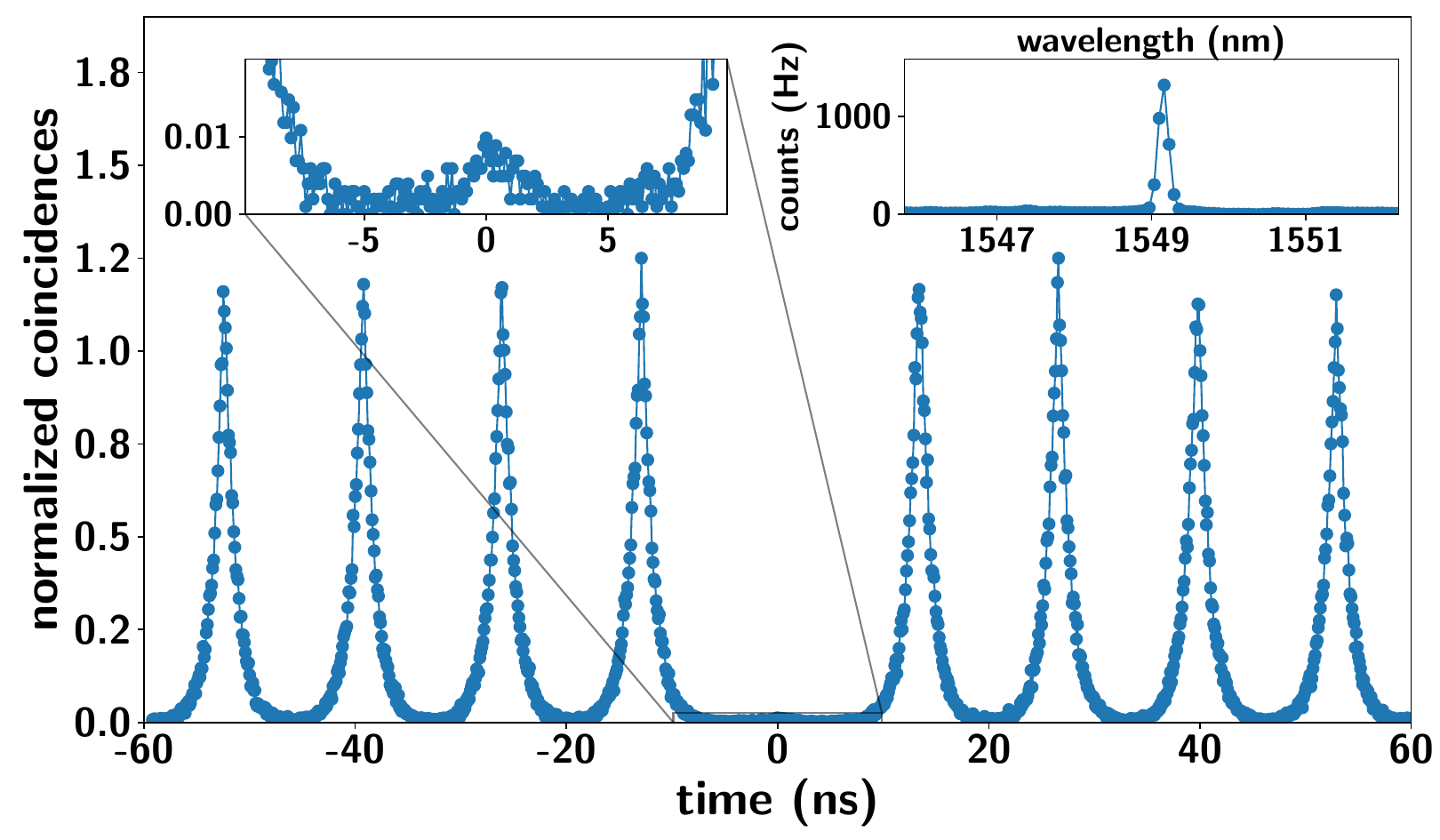}
		\caption{\textbf{Characterisation of the positively-charged exciton transition under pulsed LA excitation.} Second-order auto-correlation measurement $g^{(2)}(\tau)$ for an excitation field strength of $0.46\,\,\text{a.u.}$ and a detuning of \SI{1.5}{\nano\meter}. The well-suppressed peak at zero time delay confirms the high single-photon purity ($g^{(2)}(0)=0.018\,(1)$). Further details on the analysis of $g^{(2)}(0)$ can be found in Supplementary~Note~\ref{sec:dataAna}. The inset shows a micro photoluminescence ($\mu$-PL) spectrum of the studied transition, including spectral suppression of the laser with an excitation field strength of $0.89\,\,\text{a.u.}$ and a detuning of \SI{1.5}{\nano\meter} from the QD resonance.}
		\label{fig:characterization}
	\end{center}
\end{figure}

To start investigating and optimizing our excitation parameters, it is important to note that most quantum-secured applications rely on few trusted parameters that are typically not all experimentally accessible.
Here, we infer the photon-number probabilities $\{p_k\}$ from two measurements, the brightness ${B=\sum_{k=1}^\infty p_k}$ and the single-photon purity ${P=1-g^{(2)}(0)}$, where $g^{(2)}(0)$ is the second-order auto-correlation measurement evaluated at zero time delay.
We use an InAs QD based on an InGaAs metamorphic buffer layer enabling emission in the telecommunication C-band~\cite{Sittig2022}.
The tunable excitation is provided by a mode-locked fiber laser with a pulse length of $\SI{17(1)}{\pico\second}$ and a FWHM spectral width of $\SI{210(20)}{\pico\meter}$.
The QD transition line is filtered by a set of volume Bragg grating filters ($\text{FWHM}=\SI{0.2}{\nano\meter}$).
The total setup efficiency is determined to be \SI{13}{\percent}.
For more experimental details, see Supplementary~Note~\ref{sec:setup}.
The highest single-photon purity under pulsed LA excitation was measured as $P=0.982$, the corresponding second-order auto-correlation measurement is shown in Fig.~\ref{fig:characterization}.

\begin{figure*}
	\begin{center}
		\includegraphics[width=180mm]{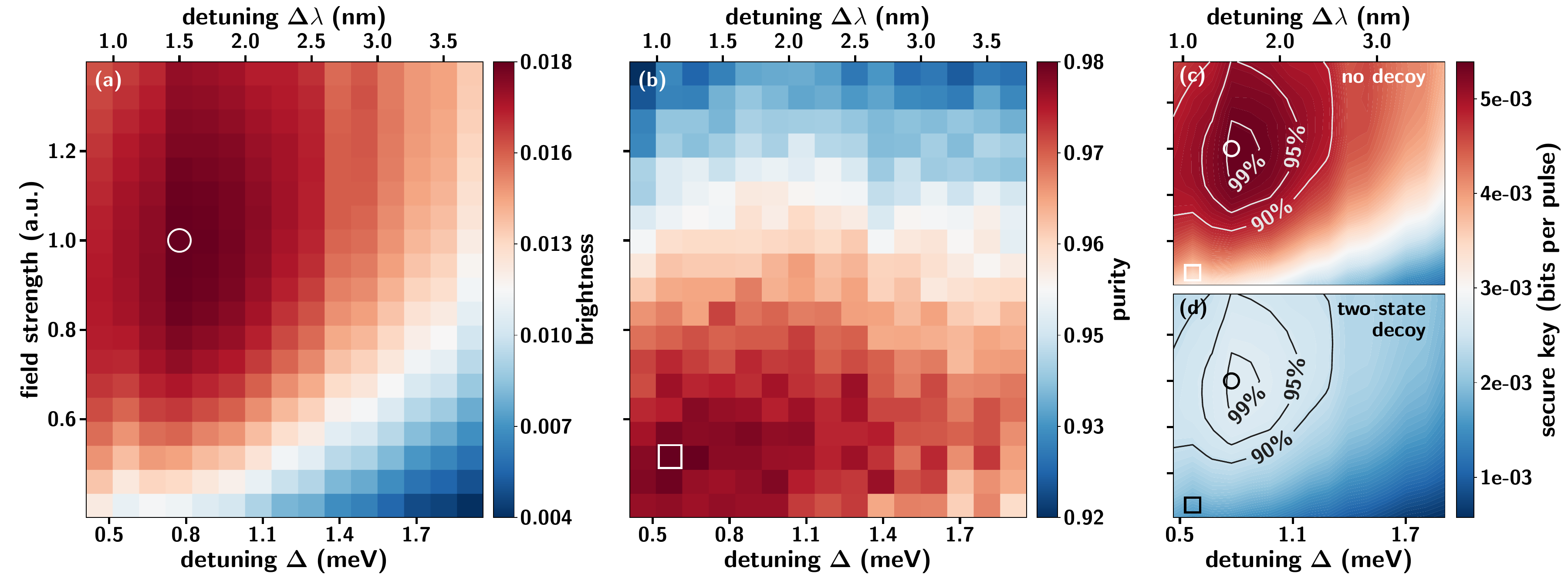}
		\caption{\textbf{Measured photon-number statistics and extrapolated QKD secure key bits per pulse for LA excitation.} Scanning the excitation parameters while simultaneously measuring (a) brightness $B=\sum_{k=1}^\infty p_k$ and (b) single-photon purity $P=1-g^{(2)}(0)$ of the QD emission. The white circle (square) marks the set of excitation parameters achieving the optimal brightness (purity). From the photon-number populations $\{p_k\}$, the secure key bits per pulse ($S\!K$) are calculated for zero distance based on the BB84 QKD protocol without (c) and with two decoy states (d). For more details on the parameter estimation see Supplementary~Note~\ref{sec:paramEst}. The equipotential lines indicate where the $S\!K$ has dropped to $\{99\,\%$, $95\,\%$, $90\,\%\}$ of their individual $S\!K$ maxima. The $S\!K$ was estimated in the asymptotic limit~\cite{GLLP04}, $SK=\eta_{\text{sif}}[Q_1(1-H_2(E_1))-f(E)Q_{\text{tot}}H_2(E_{\text{tot}})]$, where $H_2$ is the binary Shannon entropy and $\eta_{\text{sif}}=1/2$. Extrapolation for two-state decoy includes an intensity modulator loss of \SI{3}{\decibel}. Parameters for all plots are: single-photon detection error $e_{\text{d}}=0.02$, detection efficiency $\eta_{\text{d}}=0.86$, dark-count probability $Y_0=1.6\cdot10^{-6}$, error-correction code inefficiency $f=1.2$.} 
		\label{fig:exp_maps}
	\end{center}
\end{figure*}

While scanning both the power and wavelength of the pump laser, we simultaneously measure the brightness $B$ (experimentally evaluated according to Eq.~S2 and corresponding to first-lens brightness) and single-photon purity $P$. We then compile the results in 2D maps as shown in Fig.~\ref{fig:exp_maps} (a) and (b), respectively. 
Due to the low phonon density at a sample temperature of $\sim\,$\SI{4}{\kelvin} we excite the QD only with positive detunings $\Delta = \hbar(\omega_{\text{laser}}-\omega_{\text{dot}}) > 0$.
The brightness map features a single, broad maximum around $\Delta \approx \SI{0.8}{\milli\electronvolt}$($\Delta\lambda \approx \SI{1.5}{\nano\meter}$) agreeing with similar experimental findings~\cite{QBL:PRL2015, Bounouar2015} and theoretical studies~\cite{Glaessl2013Pro, Gustin.2019}.
LA excitation with sufficiently smooth pulses~\cite{Barth2016b} achieves a population inversion of the QD ground and excited state if the effective Rabi splitting of the laser-dressed states, $\hbar\Omega_{\text{eff}} = \sqrt{(\hbar\Omega)^2+\Delta^2}$, ensures an efficient exciton-phonon coupling that is characterized by the spectral phonon density $J(\omega)$~\cite{Glaessl2013Pro, QBL:PRL2015}.
The robustness of this scheme against power and wavelength fluctuations of the excitation laser is demonstrated by the broad maximum of the brightness in  Fig.~\ref{fig:exp_maps} (a) and stems from the spectral width of $J(\omega)$.
Thus, the large bandwidth of the phonon interaction directly benefits a stable operation of the QD source.
Only for large detunings and weak fields, the phonon-induced relaxation to the exciton level fails and the brightness drops significantly.
Similarly, for high powers, the effective Rabi splitting is no longer in resonance with the phonon interaction resulting in reduced brightness.

Besides emission efficiency, the single-photon purity of the quantum-light source is crucial to the performance of cryptographic protocols~\cite{BVC:npjQI22}.
Therefore, we analyze the purity $P$, depicted in Fig.~\ref{fig:exp_maps} (b), for the same parameter range as the brightness.
We identify a broad region of high purity at similar detunings but shifted towards lower powers.
At large detunings, the purity degrades because the exciton state preparation via LA phonons becomes less efficient (evident by the low brightness in the same area of Fig.~\ref{fig:exp_maps} (a)) and spurious contributions to the emission, including neighboring QDs or a quasi-continuum of transitions, are no longer negligible.
Considering a perfect two-level system, Ref.~\cite{Axt:PRL19} predicts an enhanced purity for increasing excitation field strength because the phonon-induced level inversion is delayed until the end of the pulse.
As a consequence, the chance of a reexcitation event during the same pulse, as it is known for resonant pumping~\cite{Hanschke.2018, PhysRevLett.126.233601}, would be reduced.
In our experiment, however, this process competes with, and is eventually out-weighed by, the aforementioned unintended emission decreasing the purity at high powers significantly. 

Interestingly, our experimental findings imply the absence of a trivial set of optimal parameters (simultaneously maximizing brightness and single-photon purity), which confirms some of the complex behaviours predicted in previous theory works~\cite{Axt:PRL19,BVC:npjQI22}.
Instead, a careful tuning of the excitation parameters is required for each quantum-cryptographic application.
Depending on the desired security of merit, the correct weighting of the photon-number populations $\{p_k\}$ used for the optimization~\cite{BVC:npjQI22} must be defined.
At the same time, fluctuations in the excitation parameters produce only small changes in photon-number populations.
Furthermore, optimal brightness and near-optimal purity are achieved for a pump pulse detuned by $\approx$ 1.5 nm from the QD transition that can be readily separated from the single-photon emission using efficient, off-the-shelf spectral filters.
This simplifies source operation and optimizes brightness by removing the need for a cross-polarization setup, further underlining the practicality of LA excitation for network applications~\cite{Pan:RevMod20,BAL:npjqi17,BBR:PRL18,NYC:NatComm23,BBB:NC11,PJ+:natcomm14,BOV:npj18,GAA:pra18,Kent:npjQI22,NJC:NatComms12,Zbind:PRL13,Pan:PRL14}.

We now experimentally show how to perform the excitation parameter optimization for the example of QKD, arguably the best-known primitive in quantum communication. QKD allows two parties to establish a secret key over an eavesdropped channel~\cite{Pan:RevMod20,BB84}. In that sense, the most natural figure of merit is the number of secure bits communicated per round of the protocol. This quantity can be computed from two experimental parameters: the total gain $Q_{\text{tot}}$, corresponding to the probability of detecting at least one photon from a given pulse sent by Alice, and the total error rate $E_{\text{tot}}$, indicating the fraction of states for which the wrong (polarization) detector clicks. Naturally, only the error-free single photon states contribute positively to the secure key, while the multi-photon contribution $p_{\text{m}}$ leaks significant amounts of information. Starting from experimental data, one therefore needs to estimate the values of the single-photon gain $Q_1$ and the single-photon error rate $E_1$, which are not directly accessible. In Supplementary~Note~\ref{sec:paramEst}, we infer these quantities in two ways: first by deriving an upper bound on the multi-photon emission probability
\begin{equation}
    p_{\text{m}}\leq\frac{1-Bg^{(2)}(0)-\sqrt{1-2Bg^{(2)}(0)}}{g^{(2)}(0)}
    \label{Eq:pm_bound}
\end{equation}
and second by employing the two-state decoy approach~\cite{LMC:PRL05,W:PRL05}. Compared to previous work~\cite{Waks:PRA02}, Eq.~\ref{Eq:pm_bound} gives an explicit expression for $p_{\text{m}}$ relying only on the experimentally accessible $B$ and $g^{(2)}(0)$ and provides additional intuition to Ref.~\cite{Gruenwald:19}.


\begin{figure*}
	\begin{center}
		\includegraphics[width=180mm]{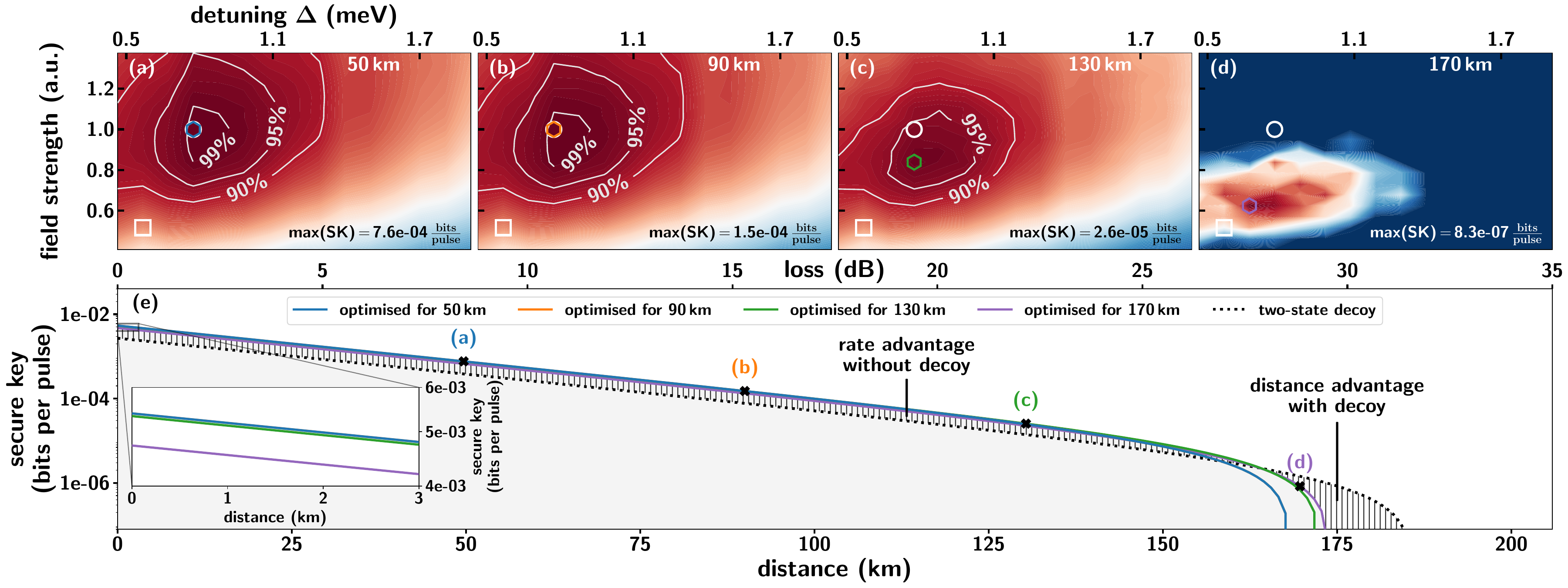}
        \caption{\textbf{Secure key rates of BB84 QKD for varying communication distances.} Secure key bits per pulse for the LA excitation parameter space in a standard BB84 QKD scenario for increasing channel length \{$50\,$km, $90\,$km, $130\,$km, $170\,$km\} (a)-(d), where we assumed a fiber attenuation of $\alpha=0.17\,$dB/km, typical for the telecom C-band. The white circle (square) marks the set of excitation parameters achieving the optimal brightness (purity) as shown in Fig.~\ref{fig:exp_maps}, whereas the colored hexagon marks the trade-off between the two, optimizing the $S\!K$ at the given distance. The color scale of each map is normalized to its maximum $S\!K$ that is noted in the bottom right corner of each map.
        (e) Calculating the $S\!K$ for each highlighted parameter set from (a)-(d) as a function transmission loss demonstrates how the tunability of LA excitation helps to adapt the emission statistics to the channel. The two-state decoy protocol reduces the $S\!K$ by a factor of $\sim$\,3 at short and medium distances but performs better in the high-loss regime. The parameters used to calculate the $S\!K$ are the same as for Fig.~\ref{fig:exp_maps}.}
		\label{fig:distance}
	\end{center}
\end{figure*}

Following the parameter estimation, we calculate the attainable secure key bits per pulse ($S\!K$) in the asymptotic regime for standard and decoy-state BB84 QKD for each set of excitation parameters as shown for zero communication distance in Fig.~\ref{fig:exp_maps} (c) and (d), respectively.
For the decoy protocol, we include a typical \SI{3}{\decibel} loss for a high-bandwidth intensity modulator required to produce the decoys.
While the qualitative dependence of the $S\!K$ is very similar for both protocols, the performance gap is evident in the absolute values.
Decoy states have been introduced to handle the risk of multi-photon contributions $p_{\text{m}}$, but since these are inherently small for QDs, introducing the constant loss of the intensity modulator outweighs the effect of an exact bounding of $p_{\text{m}}$.
Furthermore, recalling that quantum cryptography with off-resonantly or two-photon excited QDs does not require any modulator for phase scrambling, adding an intensity modulator for decoy would increase the setup complexity.
Comparing Fig.~\ref{fig:exp_maps} finally shows that for zero distance the brightness (more accurately, $p_1$) dominates the $S\!K$ map making a tight bounding of $p_{\text{m}}$ even less relevant.

However, the impact of the multi-photon events on the $S\!K$ comes into play for non-zero communication distances making the ideal set of $\{p_k\}$ no longer trivial but dependent on the channel loss.
Computing $S\!K$ maps at four distances, as depicted in Fig.~\ref{fig:distance} (a)-(d), visualizes the shift in source requirements.
Short-distance transmissions benefit most from a bright source, whereas high-loss scenarios such as long-distance communication call for sources with high purity.
Fig.~\ref{fig:distance} (e) then shows how these four ideal parameter sets behave over distances.
The difference in performance underlines the potential of individually adjusting the excitation conditions with respect to the channel loss.
Note that the joint optimization of $\{p_k\}$ by tuning the pump conditions is possible with resonant or two-photon excitation but less performant.
In Supplementary~Note~\ref{sec:FinKey}, we also present the optimal finite-size $S\!K$ for various block sizes.

The maximum distance for which the generation of a secure key is still possible is of great interest for applications. Since there is no analytical expression, we
state the maximal communication distance as minimal channel transmission $\eta_{\text{ch}}^{\text{min}}$ and find that the approximation
\begin{equation}
    \eta_{\text{ch}}^{\text{min}} \approx \frac{Bg^{(2)}(0)}{2} + Y_0\,,
    \label{eq:RuleOfThumb}
\end{equation}
where $Y_0$ is the dark-count probability, captures the break-down of secure key generation due to multi-photon contributions well under realistic assumptions.
Due to its construction (effectively lower bounding $\eta_{\text{ch}}^{\text{min}}$), Eq.~\ref{eq:RuleOfThumb} always overestimates the distance by $\sim30\,$km (see Supplementary~Note~\ref{sec:ExtendendAnalysis}).
Eq.~\ref{eq:RuleOfThumb} also implies that, within the limits of the approximation, a brighter source reduces the maximum communication distance. 
Counter-intuitive at first sight, this is readily explained as the multi-photon probability $p_{\text{m}}$ increases with the source brightness if $g^{(2)}(0)$ is unchanged (see Eq.~\ref{Eq:pm_bound}).
While brighter QDs further improve the $S\!K$ at short to medium distances, one must reduce the multi-photon component when communicating over large distances.
For this purpose, simply attenuating the signal before launching it into the untrusted channel is sufficient~\cite{Waks:PRA02}.
Note how this approach resembles the mean-photon number optimization used for QKD with WCS~\cite{Ma:Thesis08}.
Considering a detector dark-count probability $Y_0=10^{-7}$, single-photon detection error $e_{\text{d}}=0.02$ and a highly pure source ($g^{(2)}(0)=0.02$), our numerical analysis (see Supplementary~Note~\ref{sec:ExtendendAnalysis}) identifies the ideal brightness for maximum distance as $B\approx0.9\,\%$.
This is well within range of today's telecom C-band QD-technology.

Finally, we remark that implementing decoy states could be advantageous in the long-distance regime even for highly pure single-photon sources such as QDs, as reflected in Fig.~\ref{fig:distance} (e).
However, for low to moderate loss, standard BB84 outperforms the decoy-state protocol.


In conclusion, we have investigated the benefits that phonon-assisted excitation of a telecom C-band QD provides for quantum-secured applications.
Besides the convenient wavelength for communication applications, the InAs QDs provide a deep confinement potential, typically spanning over several hundred millielectron volts.
As a consequence, their photon-number statistics are relatively insensitive to temperature fluctuations~\cite{Carmesin2018}.
Moreover, the source can be operated at \SI{25}{\kelvin}, which is feasible for a low-cost Stirling cryostat~\cite{Nawrath2023}.

In addition to the previously simulated low photon-number coherence~\cite{BVC:npjQI22}, the robustness to environmental fluctuations~\cite{Armando:ACS17} and the efficient single-photon filtering, we have shown that LA excitation allows to effectively optimize the photon-number statistics with respect to the desired application.
This feature originates from interaction with the phonon environment and is therefore not common to resonant excitation schemes but can be exploited by tailoring the LA pumping conditions.
The complex implications of phonon interactions for brightness and single-photon purity have also been theoretically predicted for idealized systems~\cite{Axt:PRL19}.
Therefore, our observations can be generalized to other QD-based sources. 

As a means of improving the emission statistics independently of the excitation mechanism, temporal filtering of the signal was proposed~\cite{Ates2013, Kupko.2020} but requires special hardware and comes at the price of additional loss.
Moreover, we show in Supplementary~Note~\ref{sec:timeFilter} that optimizing $\{p_k\}$ using only the LA pump power is as efficient as temporal filtering with a fast and lossless modulator.
Only at very large distances, temporal filtering performs better since it also reduces the source brightness and detector dark counts. 

However, we note that two-photon excitation can simultaneously yield a higher brightness and purity than achievable for any parameter set using the LA scheme~\cite{Hanschke.2018}.
Nevertheless, two-photon excitation -- being a resonant process -- is sensitive to environmental fluctuations and thus less suitable for real-world implementations.

Furthermore, we found that even for quantum light sources with inherently low multi-photon contribution, decoy states can push the maximum attainable distance in QKD.
Although, in consideration of the low $S\!K$ at these distances and the experimental overhead involved, we believe that decoy states are not beneficial for QD implementations.

Finally, we would like to stress that we optimized the photon-number statistics in LA excitation with respect to QKD as an example but the process is transferable to other quantum-secured applications \cite{NYC:NatComm23,BBB:NC11,PJ+:natcomm14,SKS:NatComm23,Kent:npjQI22,GAA:pra18,NJC:NatComms12,Zbind:PRL13,Pan:PRL14} and prone to improve their performance.

\section*{Acknowledgments}

We thank R. Joos for fruitful discussions. This research was funded in whole, or in part, from the European Union’s Horizon 2020 and Horizon Europe research and innovation programme under grant agreement No 899368 (EPIQUS), the Marie Skłodowska-Curie grant agreement No 956071 (AppQInfo), and the QuantERA II Programme under Grant Agreement No 101017733(PhoMemtor); from the Austrian Science Fund (FWF) through [F7113] (BeyondC), and [FG5] (Research Group 5); from the Austrian Federal Ministry for Digital and Economic Affairs, the National Foundation for Research, Technology and Development, the Christian Doppler Research Association, the German Federal Ministry of Education and Research (BMBF) via the project QR.X (No.16KISQ013) and the European Union's Horizon 2020 research and innovation program under Grant Agreement No. 899814 (Qurope). Furthermore, this project (20FUN05 SEQUME) has received funding from the EMPIR programme co-financed by the Participating States and from the European Union's Horizon 2020 research and innovation programme.


\begin{appendices}
\setcounter{section}{0}
\renewcommand{\thesection}{\arabic{section}}
\renewcommand{\theHsection}{\arabic{section}}
\renewcommand{\thesubsection}{\Alph{subsection}.}
\setcounter{figure}{0}
\renewcommand{\thefigure}{S\arabic{figure}}
\renewcommand{\theHfigure}{S\arabic{figure}}
\setcounter{equation}{0}

\renewcommand{\theequation}{S\arabic{equation}}
\renewcommand{\theHequation}{S\arabic{equation}}

\makeatletter
\def\@seccntformat#1{\@ifundefined{#1@cntformat}%
   {\csname the#1\endcsname\space}
   {\csname #1@cntformat\endcsname}}
\newcommand\section@cntformat{SUPPLEMENTARY NOTE \thesection:\space} 
\makeatother

\section{\label{sec:setup}Experimental Setup}

\begin{figure*}
	\begin{center}
		 \includegraphics[width=180mm]{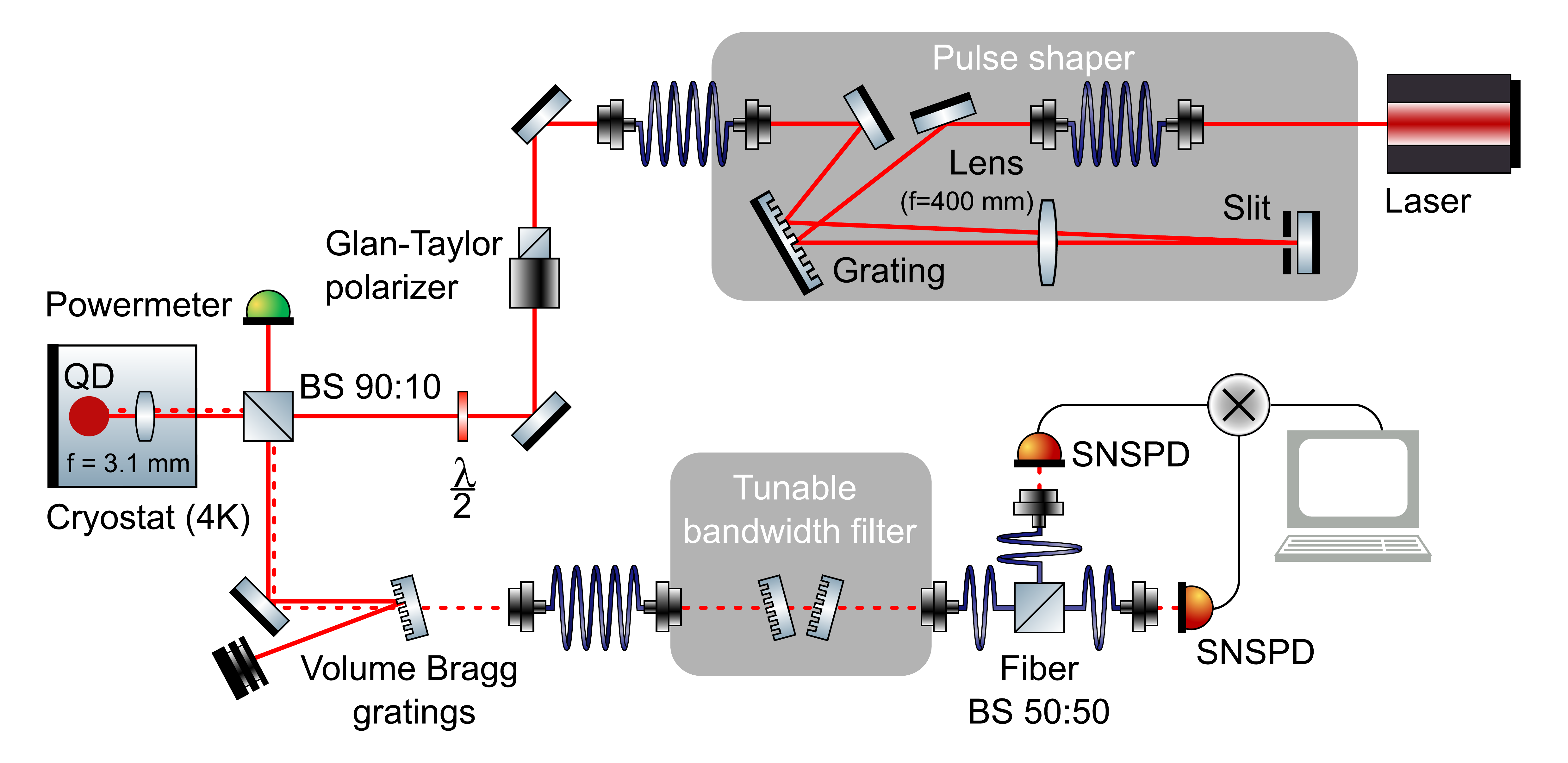}
		\caption{\textbf{Scheme of the experimental setup.} A schematic representation of the main building blocks of the experimental setup, such as excitation pulse laser, pulse shaper, excitation of the QD sample, filtering of the QD transition spectral line and detection. For more detailed setup description see the Supplementary~Note~\ref{sec:setup}.}
		\label{fig:setup}
	\end{center}
\end{figure*}

The scheme of the experimental setup is displayed in Fig.~\ref{fig:setup}. We use an Er-doped fiber pulsed mode-locked laser at a repetition rate of $\nu_{\text{rep}}=\SI{75.95}{\MHz}$ to excite the quantum dot (QD).
A filter with $\SI{1}{\nano\meter}$ bandwidth inside the laser cavity stretches the generated pulses to a pulse width of $\SI{10.1}{\pico\second}$ (spectral width FWHM $\SI{400}{\pico\meter}$, the pulses are not Fourier-limited).
The laser provides tunability of wavelength between $\SI{1530}{\nano\meter}$ and $\SI{1550}{\nano\meter}$ at an average output power $\SI{200}{mW}$.
The laser pulses are then stretched by a free space pulse shaper in $4\text{-}f$, which is based on a reflective grating (1200 lines/mm, blase at 1550 nm) with efficiency $\approx 90\,\%$, C-coated lens with a focal length of $\SI{400}{\milli\meter}$ and tunable filtering slit.
The pulses after the pulse shaper have a pulse width of $\SI{17(1)}{\pico\second}$. The pulse-shaped excitation laser beam is coupled to single-mode fiber and then collimated by an $\SI{8}{\milli\meter}$ lens collimator.
A Glan-Taylor polarizer sets the excitation beam to linear polarization and the angle of the polarization is then adjustable by a half-wave plate.
Just before the cryostat chamber a 90:10 beam splitter cube (BS) is placed to separate the incoming excitation beam and single photons emitted by the QD.
Approximately $90\,\%$ of the excitation beam (depending in its polarization state) is reflected by the BS to a power meter, which is used to control the QD excitation power, only $\approx 10\,\%$ off the laser power is guided to the cryostat chamber, where the QD sample is placed.

\begin{figure}[!h]
	\begin{center}
		 \includegraphics[width=80mm]{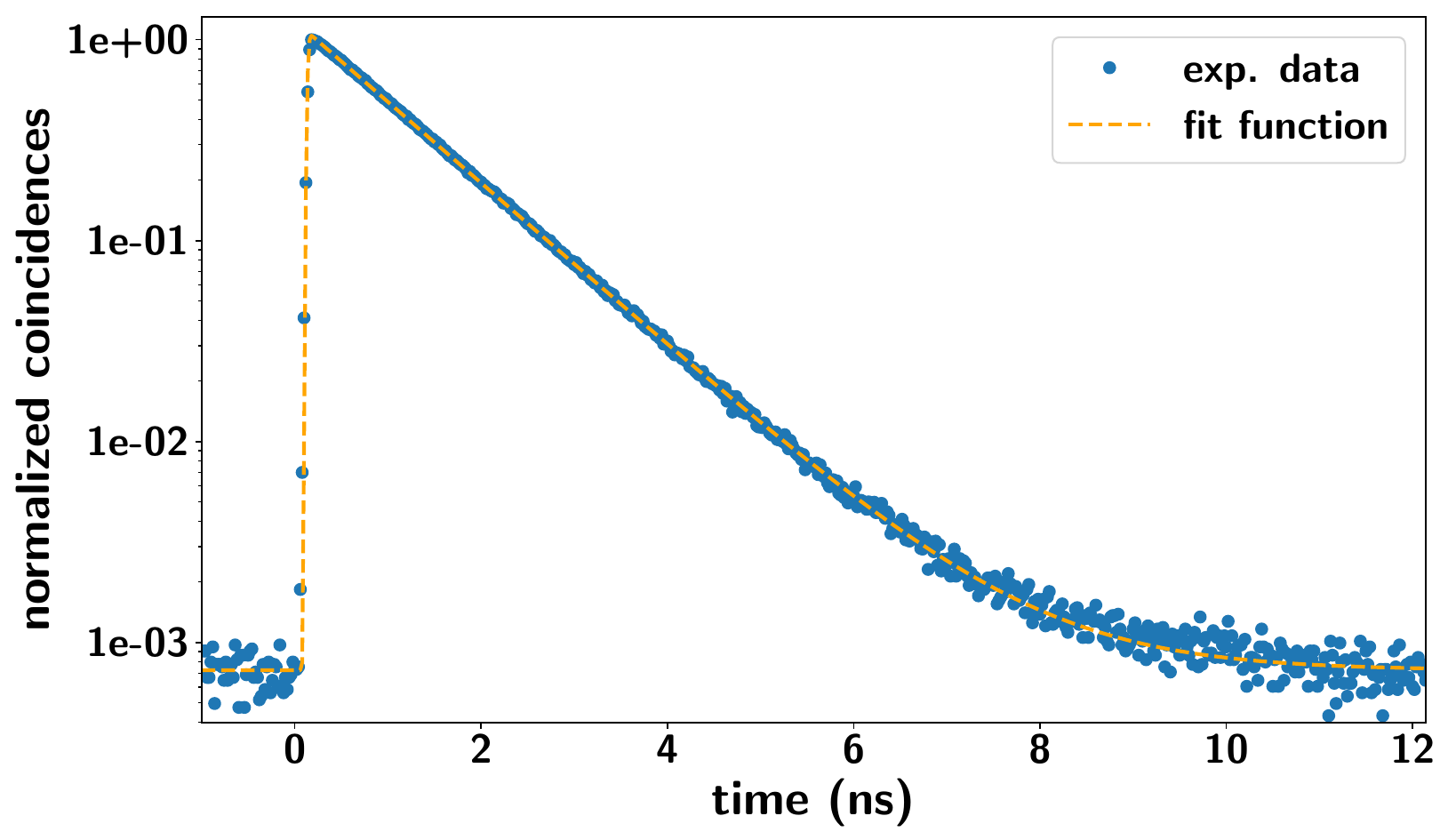}
		\caption{\textbf{Time-resolved fluorescence measurement.} For this measurement, the excitation laser was spectrally detunend by \SI{1.5}{\nano\meter} and a pump power of $0.89\,\,\text{a.u.}$ was used. The mono-exponential fit function (dashed line) yields a decay time of \SI{1.07(2)}{\nano\second}.}
		\label{fig:lifetime}
	\end{center}
\end{figure}

The sample design \cite{Sittig2022} features a bottom distributed Bragg reflector where the distance between the 23 pairs of AlAs/GaAs constituting the reflector and the semiconductor/vacuum interface corresponds to a nominal, weak $\lambda$ cavity, and the QD layer is situated in its anti-node.
The attribution of the QD transition to a positive trion, is based on power- and polarization-resolved $\mu$-PL measurements, as well as previous experimental and theoretical investigations \cite{Paul2017, Carmesin2018, Sittig2022, Dusanowski2022} on similar samples.
Fig.~\ref{fig:lifetime} displays a time-resolved fluorescence measurement in a semi-logarithmic scale where an excitation power of $0.89\,\,\text{a.u.}$ and spectral detuning of \SI{1.5}{\nano\meter} was used.
The mono-exponential fit function yields a decay time of \SI{1.07(2)}{\pico\second}.

The emission is collected by a lens ($f=\SI{3.1}{\milli\meter}$) with an NA of $0.68$. The excitation laser suppression is realized on the basis of two volume Bragg grating (VBG) notch filters with blocking a spectral bandwidth (FWHM) $\SI{1.2}{\nano\meter}$ and individual suppression OD$6$.
The QD emission is coupled by a $f=\SI{8}{\milli\meter}$ lens collimator. For precise filtering of the QD transition line we use a fiber-coupled bandwidth tunable filter based on VBGs with set filtering spectral bandwidth (FWHM) $\SI{150}{\pico\meter}$ (approximately the QD-transition linewidth).

The full setup exhibits an efficiency of $\eta_{\text{setup}}=0.13$. The measurements of $g^{(2)}(\tau)$ are acquired with a fiber-based, symmetric beam splitter in a Hanbury-Brown and Twiss configuration using superconducting nanowire single-photon detectors (SNSPDs) with an efficiency of $\eta_{\text{d}}=0.86$ each and a time tagging device.
The detection exhibits a temporal resolution (FWHM of the system response function) of $\SI{34}{\pico\second}$. The dark counts of the used detectors are \SI{130}{\hertz} for the first detector and \SI{180}{\hertz} for the second one.

\section{\label{sec:dataAna}Evaluation of Time Tags}

For each laser detuning and pump power, a single measurement run is performed from which both the brightness $B$ and purity $P$ are inferred.
To obtain enough statistics in the auto-correlation data, each measurement is stopped once the coincidence counts of the uncorrelated side peaks exceed a threshold value (here, $700$\,counts at \SI{100}{\pico\second} bin width).

Simply summing the count rates of both detection channels results in probabilistic double-counting of multi-photon states
\begin{equation}
    \Tilde{B}=B + \sum_{k\geqslant2}p_k\left(1-\frac{1}{2^{k-1}}\right) = B + \frac{1}{2}p_2  + \frac{3}{4}p_3\,.
\end{equation}

Thus, to avoid over-counting we subtract all coincidence events occurring within one repetition period $T_{\text{rep}}=\tfrac{1}{\nu_{\text{rep}}}$ such that the measured brightness reads as
\begin{equation}
    \label{eq:S_B_exp}
    B=\frac{T_{\text{rep}}}{\eta_{\text{setup}}\eta_{\text{d}}}\times\Big(R_1 + R_2 - CC_{1,2}(t_{\text{coinc}}=T_{\text{rep}}) \Big) 
\end{equation}
where $R_i$ is the raw count rate of the $i-$th detector and $CC$ denotes the coincidence count rate. 

The auto-correlation measurements (compare Fig.1 in the main text) use a bin width of \SI{100}{\pico\second} and are numerically evaluated to deduce $g^{(2)}(0)$. 
We only apply a background subtraction based on the expected coincidences arising from a dark count in at least one of the channels computed as
\begin{equation}
    CC_{1,2}^{\text{d}} = R_1R_2^{\text{d}} + R_2R_1^{\text{d}} + R_1^{\text{d}}R_2^{\text{d}}
\end{equation}
where the superscript 'd' denotes the dark count rates.

Integrating the area over the central repetition period and dividing by the averaged and blinking-corrected area of the outer peaks then yields the $g^{(2)}(0)$ value.
For applications in quantum communication, it is crucial to consider the full repetition period when calculating the source's purity since an adversary has access to all the information leaving the sender's lab.

\section{\label{sec:paramEst}Parameter Estimation for Quantum Key Distribution with Quantum Dots}

\subsection{Practical asymptotic secure key rate}

When assessing the performance of a QKD network, it is necessary to estimate the fraction of securely exchanged qubits, or $untagged$, in order to separate them from the $tagged$ qubits where some information could have been leaked.
This step, called parameter estimation, will determine the amount of necessary privacy amplification and will therefore be crucial to guarantee practical information-theoretic security. In discrete-variable photonic implementations, the most widely used security proofs assume some form of active (or passive) phase randomization~\cite{LP:QIC07}, in order to separate the contributions from different photon number components. For LA-excited QDs, it is a fair assumption that the emitted photon states bear very little coherence between the photon-number states~\cite{BVC:npjQI22}. We may then proceeed assuming that only the single photons states contribute positively to the secure-key generation, whereas multi-photon states carry redundantly encoded information which could be extracted with Photon Number Splitting (PNS) attacks for instance~\cite{BLM:PRL00}.

We start by briefly recalling some relevant quantities in a practical BB84 QKD scenario.
Let us define the yield of a $k$-photon state as the conditional probability of a detection on the receiver's detector given that the sender generates a $k$-photon state:
\begin{equation}
\label{eq:S_Yk}
Y_k = Y_0 + (1 - Y_0)\left[1 - (1 - \eta_\text{d} \eta_\text{ch})^k\right] \ , 
\end{equation}
where $\eta_\text{d}$ is the detection efficiency, $\eta_\text{ch}$ is the channel transmission, and $Y_0$ is the dark-count probability.
We define the gain $Q_k$ of a $k$-photon state as the probability of a detection event resulting from this state
\begin{equation}
Q_k = p_k  Y_k \ . 
\end{equation}
where $p_k$ is the probability of $k$-photon emission from the QD source.
Further, we define $e_k$, the error rate of the $k$-photon state, as
\begin{equation}
\label{eq:S_ek}
e_k = \frac{e_0 Y_0 + e_\text{d} \left[1 - (1 - \eta_\text{d} \eta_\text{ch})^k\right]}{Y_k} \ , 
\end{equation}
where the parameter $e_\text{d}$ characterizes the detection error probability, dependent on the optical alignment of the entire system, and $e_0$ is the error rate of the background, which, if we assume to be random, is $e_0 = \frac{1}{2}$.
In a QKD implementation, the receiver measures the total gain of the signal state $Q_{\text{tot}}$
\begin{equation}
    \label{eq:S_Qtot}
    Q_{\text{tot}} = \sum_{k=0}^\infty Q_k
\end{equation}
 and the qubit error rate $e_{\text{tot}}$
\begin{equation}
    \label{eq:S_Etot}
    e_{\text{tot}} = \frac{1}{Q_{\text{tot}}} \sum_{k=0}^\infty e_k Q_k
\end{equation}
and, after estimating the single-photon gain $Q_1$ and error $e_1$, computes the rate of secure key bits per pulse ($S\!K$) with the GLLP formula \cite{GLLP04}
\begin{equation}
    \label{eq:S_GLLP}
    S\!K = \frac{1}{2}[Q_1 (1 - H_2(e_1)) - f(e_{\text{tot}})Q_{\text{tot}} H_2(e_{\text{tot}})] \ .
\end{equation}
In this formula, the term $f(e_{\text{tot}})Q_{\text{tot}} H_2(e_{\text{tot}})$ accounts for the cost of error correction - $f(e_{\text{tot}})$ being the error correcting code inefficiency and $H_2(\cdot)$ the binary entropy - and $Q_1 (1 - H_2(e_1))$ states that only the error-free single photon states contribute to the secure key generation.

Eqs.~\ref{eq:S_Yk} and \ref{eq:S_ek} describe theoretical values, thus we will now describe two different procedures to estimate $e_1$ and $Q_1$ from experimentally accessible quantities.

\subsection{Estimation of single-photon parameters based on auto-correlation functions}
We will estimate the multi-photon contribution relying only on the brightness $B$ and single photon purity $P$. Such quantities are indeed the main source parameters, and are readily measured by the sender.
However, since the channel parameters are untrusted, the honest parties have to assume that all losses and errors arise from single photon states, that is $Y_k = 1 $ and $ e_k = 0 $ for $k \geqslant 2$. 

Thus, the single photon parameters can be estimated as follows:
\begin{equation}
    \label{eq:S_Q1}
    Q_1 \geqslant  Q_{\text{tot}} - p_\text{m} - Y_0 p_0
\end{equation}
\begin{equation}
    \label{eq:S_e1}
    e_1 \leqslant \frac{e_{\text{tot}}Q_{\text{tot}} - \frac{1}{2} Y_0 p_0}{Q_1} \ ,
\end{equation}
where $p_0 = 1 - B$ and $p_\text{m} = \sum_{k\geqslant 2} p_k$ is the multi-photon probability. Note that, implicit in the above equations, is the assumption that parties have a trusted estimation of the vacuum contribution. We bound $p_\text{m}$ starting from the second-order auto-correlation function, along with the reasonable assumptions that $p_{n\geqslant 4} = 0$  and $p_1 > p_{\text{m}}$,  

\begin{align}
\label{eq:g2_bound}
    g^{(2)}(0) = \frac{2p_2 + 6p_3}{(p_1 + 2p_2 + 3p_3)^2} = \frac{2p_\text{m} + 4p_3}{(p_1 + 2p_\text{m})^2} \nonumber \\ \times \left[ 1 + \frac{p_3}{p_1 + 2p_\text{m}} \right]^{-2} \simeq \frac{2p_\text{m}}{(p_1 + 2p_\text{m})^2}\ ,    
\end{align}

and then truncate it at the zeroth order in $p_3$.
Computing the error $F(p_1, p_2, p_3)$ introduced by our approximations on the $ g^{(2)}(0)$, we note that $F(p_1, p_2, p_3) > 0$ for all $ p_k \in (0, 1)$ with  $k=1,2,3$ -- proving that the right-hand side of Eq.~\ref{eq:g2_bound} provides an actual lower bound -- and $F(p_1, p_2, 0) = 0$ showing that it holds tight in the limit of vanishing $p_3$.
After rewriting Eq.~\ref{eq:g2_bound} as a lower bound and expressing it in terms of brightness $B$,
\begin{equation}
\label{eq:g2_bound_B}
    g^{(2)}(0) \ge \frac{2p_\text{m}}{(B+p_\text{m})^2} \ ,
\end{equation}
we expand and rearrange again
\begin{equation}
\label{eq:g2_bound_B_expanded}
   p_\text{m}^2 + \left(2B-\frac{2}{g^{(2)}(0)}\right)p_\text{m} + B^2 \ge 0 .
\end{equation}
Since $p_\text{m}\ge 0$ and $1\ge B\ge0$, the inequality has only one solution for a single-photon source (i.e.${\textstyle \frac{1}{2}}\ge g^{(2)}(0)\ge 0$~\cite{Gruenwald:19})
\begin{equation}
\label{eq:S_pm_bound}
    p_\text{m} \le \frac{1 - B g^{(2)}(0) - \sqrt{1 - 2Bg^{(2)}(0)}}{g^{(2)}(0)} \ .
\end{equation}
Eq.~\ref{eq:S_pm_bound} holds for any source with sub-Poissonian emission,  at any distance, and providing an explicit bound which only depends on experimentally accessible parameters.

\subsection{Estimation of single-photon parameters based on decoy states}
In the previous scenario, Alice and Bob only exchange signal states to establish a secure shared key.
However, this leads to an estimation of single photon parameters that is not tight (Eqs.~\ref{eq:S_Q1} and \ref{eq:S_e1}).
As a countermeasure, one can let Alice modulate the intensity of the states she sends, chosen from the set $\{\rho, \nu_1, \nu_2 ...\}$, in a way that is unknown to the eavesdropper.
After the quantum step of the protocol, Alice and Bob can evaluate the total gain and error rate for each state $\{Q_{\text{tot}}^{(i)}, e_{\text{tot}}^{(i)} \}$ and solve the system of equations for $\{Y_k, e_k\}$.

These so-called $decoy$ states have been shown to increase the achievable $S\!K$ drastically for implementations based on attenuated laser pulses \cite{LMC:PRL05}.
Even though decoy states decrease the sifting efficiency $\eta_{\text{sif}}$ as they cannot contribute to the raw key, a tighter estimation of the single-photon contribution is preferable, especially for large distances.

When working with sub-poissonian sources, the multi-photon component $p_{n\ge4}$ can be neglected and two decoy states are sufficient to compute the yields and error rates, $\{Y_k, e_k\}$, exactly. This implies that we can use the theoretical formulas in Eqs.~\ref{eq:S_Yk} and \ref{eq:S_ek}.

\section{\label{sec:FinKey}Finite key analysis}

In this section we will briefly sketch a security analysis that also accounts for finite-key effects, following \cite{Morrison2023}.
In particular, a protocol is said to be $\varepsilon_{\text{cor}}$-correct if the final key shared between Alice and Bob are identical with probability higher than $1 - \varepsilon_{\text{cor}}$, and $\varepsilon_{\text{sec}}$-secret if the information exposed to an eavesdropper, in the case where the protocol does not abort, is limited by $\varepsilon_{\text{sec}}$.
Formally, these two definitions are expressed as

\begin{gather}
    \label{eq:S_secdef}
    P\left[ K_A \ne K_B \right] \le \varepsilon_{\text{cor}} \ , \nonumber \\
    (1 - p_\text{abort}) || \rho_{A E} - M_A \otimes \rho_E||_1 \le \varepsilon_{\text{sec}} \ ,     
\end{gather}

where $K_A$ and $K_B$ are the secure keys held by the honest parties at the end of the protocol, $p_\text{abort}$ is the probability to abort the protocol, $\rho_{AE}$ is the joint classical-quantum state of the honest party and eavesdropper, and $M_A$ is the uniform mixture of all possible values of $K_A$.
In this scenario, the protocol is said to be $\varepsilon_{\text{qkd}}$-secure with $\varepsilon_{\text{qkd}} \ge  \varepsilon_{\text{cor}} + \varepsilon_{\text{sec}}$.
We further note that, in the standard implementations of the BB84 protocol, the secrecy crucially relies on the classical steps of parameter estimation $\varepsilon_{\text{PE}}$, error correction $\varepsilon_{\text{EC}}$ and privacy amplification $\varepsilon_{\text{PA}}$, for which $\varepsilon_{\text{sec}} > \varepsilon_{\text{PE}} + \varepsilon_{\text{EC}} + \varepsilon_{\text{PA}}$ must hold.

\begin{figure*}
	\begin{center}
		 \includegraphics[width=140mm]{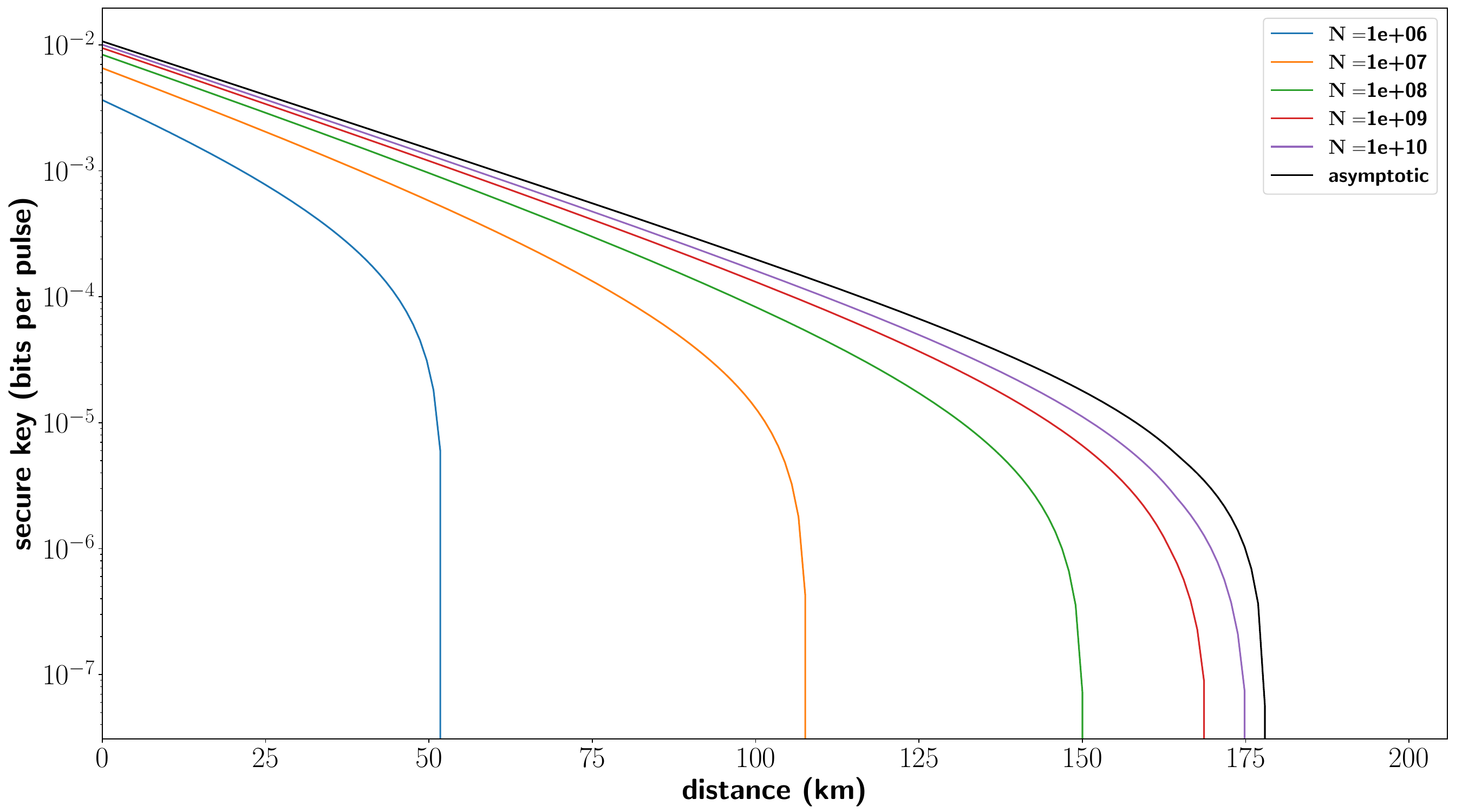}
		\caption{\textbf{Finite-key analysis for QKD.} The secure key as bits per pulse, $S\!K$, including finite-key effects calculated according to Eq.~\ref{eq:finitekeySKR} is shown for multiple block sizes. At each distance the $S\!K$ was optimized over the probability of sending the qubit in the X basis, $p_\text{X}$, and excitation conditions resulting in the ideal photon-number populations $\{p_i\}$. The parameters used for extrapolating $S\!K$ are: single-photon detection error $e_{\text{d}}=0.02$, detection efficiency $\eta_{\text{d}}=0.86$, dark-count probability $Y_0=1.6\cdot10^{-6}$ and error-correction code inefficiency $f=1.2$}
		\label{fig:finitekey}
	\end{center}
\end{figure*}

We will consider the Efficient BB84 protocol, that exploits one basis for key generation and the other for error estimation without sacrificing the security of the implementation \cite{Lo2004}.
Exploiting the $X$ basis for key and the $Z$ for error, the number of events after the information reconciliation step is $ n^{\text{b}} = N p_{\text{b}}^2 Q_{\text{tot}} $, where $p_{\text{b}}$ is the probability of choosing the basis $\text{b} = X, Z$ and $N$ is the total number of rounds performed.
We can isolate the clicks caused by non multi-photon pulses as $ n^{\text{b}}_{\text{sp}} = n^{\text{b}} - n^{\text{b}}_{\text{mp}}$ with $n^{\text{b}}_{\text{mp}} = N p_{\text{b}}^2 p_{m}$. Note that, in the asymptotic analysis (see Eqs.~\ref{eq:S_Q1} and \ref{eq:S_e1}), we subtracted the vacuum states contribution as well. However here, following \cite{Morrison2023}, we will lump together vacuum and non multi-photon component and, for the comparison in Fig.~\ref{fig:finitekey}, we adjusted the asymptotic equations accordingly.

This quantity can be lower bounded deriving an upper limit for the multi-photon contribution based on the Chernoff bound, which, for a sum of binary variables $x = \sum x_j$ with $x_j \in \{0, 1\}$, is given by $\overline{x} = (1 + \delta) x$ with $\delta = \frac{\beta + \sqrt{8\beta x + \beta^2}}{2x}$ and $\beta = - \ln(\varepsilon_{\text{PE}})$. Thus, a conservative estimation of the clicks contributing to the secure key generation reads
\begin{equation}
    \label{eq:S_n_sp_bound}
    \underline{n}^{\text{b}}_{\text{sp}} = n^{\text{b}} -  \overline{n}^{\text{b}}_{\text{mp}}  \ .
\end{equation}
Analogously, we define $ m^{\text{b}} = N p_{\text{b}}^2 e_{\text{tot}} Q_{\text{tot}}$ the total number of errors in the basis $\text{b}$ and, consequently, the bit error rate on the single photon contribution reads
\begin{equation}
    \label{eq:S_beta_sp}
    \sigma^{\text{b}} = \frac{m^{\text{b}}}{\underline{n}^{\text{b}}_{\text{sp}}} \ ,
\end{equation}
having implicitly assumed the worst-case scenario that all the errors stem from the non multi-photon events.
We recall that, in the implementation, we will compute the bit error rate on the $n^{\text{z}}$ $\phi^{\text{z}}$ bits exchanged in the $Z$ basis, that will be useful to upper bound the phase error rate in the $X$ basis as \cite{Yin2020}
\begin{equation}
    \label{eq:S_phi_sp}
    \overline{\phi}^{\text{x}} = \sigma^{\text{z}} + \gamma(n^{\text{x}}, n^{\text{z}}, \sigma^{\text{z}}, \varepsilon_{\text{PA}}) \ ,
\end{equation}
where
\begin{gather} 
    \gamma(n, k, \lambda, \varepsilon) = \frac{1}{2 + 2 \frac{A^2G}{(n+k)^2}} \nonumber \\ \times \left[ \frac{(1 - 2\lambda) AG}{n + k} +\sqrt{\frac{A^2 G^2}{(n+k)^2} + 4\lambda (1-\lambda) G} \right] \ ,    
\end{gather}

\begin{gather} 
     A = \max\{n, k\} \ , \nonumber \\ G = \frac{n + k}{nk} \ln \frac{n + k}{2 \pi n k \lambda (1 - \lambda) \varepsilon^2 } \ .   
\end{gather}

This leads to a secure key rate 
\begin{equation}
      S\!K =  \frac{1}{N} \left[\underline{n}^{\text{x}}_{\text{sp}} (1 -  H_2(\overline{\phi}^{\text{x}})) - \lambda_{\text{EC}} - 2\log_2 \frac{1}{2\varepsilon_\text{PA}} - \log_2 \frac{2}{\varepsilon_\text{cor}}\right]
    \label{eq:finitekeySKR}
\end{equation}
where $\lambda_{\text{EC}} = n^{\text{x}} f(e_{\text{tot}}) H_2(e_{\text{tot}}) $ are the bits leaked during error correction.

\section{\label{sec:ExtendendAnalysis}Extended Analysis of Secure Key Generation with Quantum Dots}

\subsection{Maximum QKD-distance approximation from experimental measures}

Estimating the maximum attainable communication distance $d_{\text{max}}$ for QKD for a given source, has high relevance for practical implementations.
However, due to the complexity of Eq.~\ref{eq:S_GLLP}, one cannot solve it analytically for the channel transmission $\eta_{\text{ch}}$ but has to resort to numerical methods when an tight approximation is required.
On the other hand, even the numerical evaluation of the $S\!K$ for a given photon-number statistics always involves estimating other protocol parameters such as single-photon detection error, detection efficiency, dark-count probability and error-correction code inefficiency. 
Therefore, an analytic approximation for $d_{\text{max}}$ can be advantageous -- especially if the required quantities are easily accessible.  

We start by upper bounding Eq.~\ref{eq:S_GLLP} by
\begin{equation}
\label{eq:S_boundSK}
    S\!K\,\leq\,\frac{1}{2}Q_1(1-H_2(e_1))
\end{equation}
as the cost for error correction,  $f(e_{\text{tot}})Q_{\text{tot}}H_2(e_{\text{tot}})$, is strictly positive.
The maximum attainable distance $d_{\text{max}}$ is formalized as the minimal channel efficiency $\eta_{\text{ch}}^{\text{min}}$ for which $S\!K>\delta$ where a threshold of $\delta=10^{-8}$ is used here.
From Eq.~\ref{eq:S_boundSK} follows $S\!K\rightarrow0$ if $\frac{1}{2}Q_1(1-H_2(e_1)) \rightarrow 0$, or further simplified $Q_1\rightarrow0$.
Note that $e_1\rightarrow\frac{1}{2}$ results in a vanishing $S\!K$ but since  $Q_1\rightarrow0$ also implies $e_1\rightarrow\frac{1}{2}$ (see Eqs.~\ref{eq:S_Qtot},~\ref{eq:S_Etot},~\ref{eq:S_Q1},~\ref{eq:S_e1}) we focus only on $Q_1\rightarrow0$.
Inserting Eq.~\ref{eq:S_Qtot} into Eq.~\ref{eq:S_Q1} and rearranging yields
\begin{equation}
    \label{eq:S_Q1explizit}
    Q_1=p_{\text{m}}\left(\frac{p_1}{p_{\text{m}}}Y_1 + Y_2 -1\right).
\end{equation}
Since $p_{\text{m}}\,\geqslant\,0$ for a realistic source, the expression inside the bracket has to tend to zero to cause $Q_1\rightarrow0$.
Assuming $Y_0\ll1$ we can simplify Eq.~\ref{eq:S_Yk} to $Y_1=\eta_{\text{ch}}-Y_0$ and $Y_2=2\eta_{\text{ch}}-\eta_{\text{ch}}^2-Y_0$ such that $Q_1\geqslant 0$ entails
\begin{equation}
    1\,\leq\,\eta_{\text{ch}}\left(\frac{p_1}{p_{\text{m}}} + 2 - \eta_{\text{ch}}\right) - Y_0\left(\frac{p_1}{\pmulti}+1\right).
\end{equation}
As we are looking for long-distance communication, we use $\eta_{\text{ch}}\ll1$ and rearrange 
\begin{equation}
    \label{eq:S_eta_ch_less}
    \eta_{\text{ch}}\,\geqslant\,\cfrac{1}{\cfrac{p_1}{p_{\text{m}}}+2} + Y_0\cfrac{\cfrac{p_1}{p_{\text{m}}}+1}{\cfrac{p_1}{p_{\text{m}}}+2}\,\geqslant\,\cfrac{1}{\cfrac{p_1}{p_{\text{m}}}+2} + Y_0\,.
\end{equation}
The maximum distance at which a secure key can still be generated now corresponds to the minimal channel transmission that satisfies Eq.~\ref{eq:S_eta_ch_less}. Therefore, we write
\begin{equation}
\label{eq:S_eta_min}
    \eta_{\text{ch}}^{\text{min}}\,\approx\,\cfrac{1}{\cfrac{p_1}{p_{\text{m}}}+2} + Y_0\,
\end{equation}
and finally, use Eq.~\ref{eq:S_pm_bound} and $p_1=B-\pmulti$ to re-express the result only in terms of $B$ and $g^{(2)}(0)$ as
\begin{equation}
    \eta_{\text{ch}}^{\text{min}}\,\approx\,\cfrac{1}{\cfrac{Bg^{(2)}(0)}{1-Bg^{(2)}(0)-\sqrt{1-2Bg^{(2)}(0)}}+1} + Y_0\,.
\end{equation}
Since $Bg^{(2)}(0)\ll1$, we can expand $Bg^{(2)}(0)$ in a Taylor series and truncate after the first order such that
\begin{equation}
    \label{eq:S_MaxDist}
    \eta_{\text{ch}}^{\text{min}}\,\approx\,\frac{Bg^{(2)}(0)}{2} + Y_0.
\end{equation}
For state-of-the-art technology, we have $Y_0\ll Bg^{(2)}(0)$ such that one can also dismiss $Y_0$ in the above equation.

Considering the complexity of Eq.~\ref{eq:S_GLLP}, the approximation is strikingly simple and follows directly from the fundamental source parameters.
Yet, for a broad parameter range, the results compare well to the numeric solution where Eq.~\ref{eq:S_MaxDist} always overestimates the maximum distance.
This systematic error is rooted in the approximation's derivation as an upper bound and is primarily caused by disregarding the error correction term of Eq.~\ref{eq:S_GLLP}.
Typically, the overestimation amounts to $25-35$km.

In the following section, we juxtapose the approximation to the numeric results for our source.

\subsection{Photon-number optimization via variable attenuation}

In Supplementary~Note~\ref{sec:paramEst} we inferred all experimental quantities required to calculate the $S\!K$ directly from the estimated photon-number populations $\{p_k\}$. To include the variable attenuation, we now model the photon loss first. To this end, we apply
\begin{equation}
    \begin{array}{lcl}
    p_0(\eta_{\text{att}}) & = & p_0 + p_1 (1-\eta_{\text{att}}) + p_{\text{m}} (1-\eta_{\text{att}})^2 \\
    p_1(\eta_{\text{att}}) & = & p_1 \eta_{\text{att}} + p_{\text{m}} (1-\eta_{\text{att}}^2-(1-\eta_{\text{att}})^2)\\
    p_{\text{m}}(\eta_{\text{att}}) & = & 1 - p_0(\eta_{\text{att}}) - p_1(\eta_{\text{att}})
    \end{array}
    \label{eq:beamsplitter}
\end{equation}
where we used $p_2\gg p_3$ and $\eta_{\text{att}}$ is the probability to transmit a photon. Note that this model is equivalent to a beam splitter with tunable reflectivity. The approach is similar to the one used in Ref.~\cite{Waks:PRA02}.

\begin{figure*}
	\begin{center}
		\includegraphics[width=180mm]{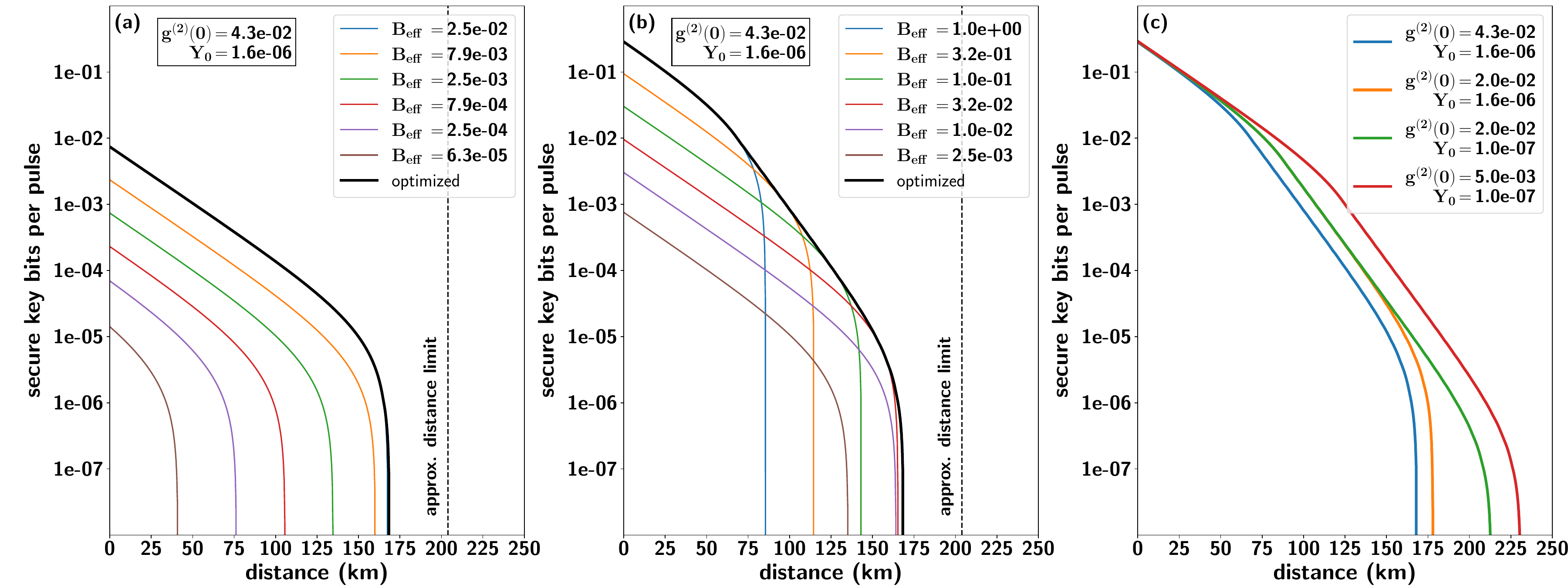}
		\caption{\textbf{Secure key bits per pulse over distance for variable signal attenuation.} 
        Including a variable attenuator into the sender's setup can improve the secure key bits per pulse ($S\!K$) at large distances. (a)-(b) The $S\!K$ as a function of distance is displayed for six effective brightness values, $B_{\text{eff}}$. The thick, black line represents the attainable $S\!K$ for a point-wise optimization of $B_{\text{eff}}$. The dotted vertical line indicates the approximated maximum distance according to Eq.~\ref{eq:S_MaxDist}. The dark-count probability $Y_0=1.6\cdot10^{-6}$ and purity of $95.7\,\%$ correspond to the experimental data at ideal brightness excitation conditions. For (a) the maximum brightness $B=2.5\,\%$ corresponds to the experimental value whereas for (b) an ideal brightness, $B=100\,\%$, is assumed. (c) Point-wise optimized curves for a set of $\{g^{(2)}(0)$, $Y_0\}$ highlighting the different influences on the $S\!K$. The parameters for all plots are: single-photon detection error $e_{\text{d}}=0.02$, detection efficiency $\eta_{\text{d}}=0.86$, error-correction code inefficiency $f=1.2$, fiber attenuation $\alpha=0.17\,$dB/km.}
		\label{fig:trustedBS}
	\end{center}
\end{figure*}

With the modified set of $\{p_k\}(\eta_{\text{att}})$, we proceed as before estimating $Q_1$, $e_1$, $Q_{\text{tot}}$, $e_{\text{tot}}$ and calculating $S\!K$.
For each communication distance, we optimize the $S\!K$ over $\eta_{\text{att}}$ to assess the full potential of a given source.
This process resembles the optimization used to identify the ideal mean-photon number in QKD with weak coherent states~\cite{Ma:Thesis08}.
The results are depicted in Fig.~\ref{fig:trustedBS} where we compare our source to idealized sources and detection.

For the experimental source parameters, we find that the maximum brightness is -- by coincidence -- very close to the point-wise optimized curve. Hence, reducing the effective brightness ($B_{\text{eff}}(\eta_{\text{att}})=p_0(\eta_{\text{att}})+p_1(\eta_{\text{att}})+p_{\text{m}}(\eta_{\text{att}})$) will reduce the maximum communication distance and the $S\!K$ at short distances.
However, assuming a brighter source, as in Fig.~\ref{fig:trustedBS}~(b), the results change drastically and the benefit of adjusting the attenuation according to the channel loss becomes clear.
The enveloping curve (i.e. point-wise optimized) now features three regions with successively larger exponential decrease in the $S\!K$.
Up to $\sim\SI{75}{\kilo\meter}$, the best $S\!K$ is achieved without any attenuation, since $S\!K$ is predominantly set by $p_1$.
The next region is shaped by the continuous balancing of $p_1(\eta_{\text{att}})$ and $p_{\text{m}}(\eta_{\text{att}})$ to optimize $S\!K$. 
Finally, at around \SI{170}{\kilo\meter}, further attenuation cannot push the maximum distance anymore as the impact of the dark counts dominates.

We indicate in Fig.~\ref{fig:trustedBS}~(a)-(b) the maximum distance approximation obtained from Eq.~\ref{eq:S_MaxDist} where the brightness corresponds to the ideal long-distance brightness. We see that Eq.~\ref{eq:S_MaxDist} overestimates the distance by $\sim30\,$km as discussed above.

To analyze the impact of distinct experimental parameters on the $S\!K$, we simulate the results for different purities and dark-count probabilities, $Y_0$, while the brightness is always assumed as $B=100\,\%$.
Only displaying the attenuation-optimized $S\!K$ for each parameter pair, we see how the position of the first inflection point is influenced by $p_{\text{m}}$ (see Eq.~\ref{eq:S_pm_bound} for fixed $B$ and changing $g^{(2)}(0)$) whereas the distance at which the $S\!K$ curve drops for the second time is determined by both, $p_{\text{m}}(\eta_{\text{att}})$ and $Y_0$.
The two inflection points can readily be associated with different causes for the $S\!K$ to break down. 
In the first case, the information leakage due to multi-photon events is too large as to permit the sifting of a secure key from the raw key while in the second case, signal clicks are similarly probable as dark-count clicks resulting in a high error probability $e_1$. 

From the above findings, we conclude that a brighter but similarly pure source, as already available in the C-band~\cite{Nawrath2023}, increases the $S\!K$ at short and medium distances but will not allow to reach greater distances.
The purity on the other hand, has little effect on the $S\!K$ for a short channel but improves $S\!K$ for medium distances and -- in combination with a low dark-count probability -- boosts the maximum attainable communication distance $d_{\text{max}}$.

\section{\label{sec:timeFilter}Time Filtering}

In the main text, we showed how tuning the excitation conditions of the LA scheme changes the photon-number populations $\{p_k\}$ of the QD source.
In a similar way, temporal filtering can be used to manipulate the photon-number statistics and improve secure key rates of QKD~\cite{Ates2013, Kupko.2020}.
This technique can be implemented irrespective of the excitation scheme but requires a fast amplitude modulator with sufficient suppression at the sender's site.
In the following, we compare the two methods based on our experimental data.

The idea of a temporal filter is to enhance the single-photon purity of the source, and at the same time, reduce the dark-count probability at the receiver.
In an experimental realization, the fast amplitude modulator is phase-locked to the driving laser and transmits only during a gating window defined by its widths $\tau_{\text{A}}$ and its delay $t_0$ to the reference input.
By adjusting $t_0$ such that the gating window starts just before the probability of a passing photons peaks (i.e. the peak in the TCSPC measurement), one can vary $\tau_{\text{A}}$ to decide how much of the exponential decay trace should be transmitted. While reducing the effective brightness, this technique usually improves the purity by excluding two-photon events caused by refilling, or by reducing the single-photon contribution from neighbouring QDs.

\subsection{Time filtering via post-selection}

In this section, we resort to time filtering by post-processing.
We, however, emphasize that a secure implementation of time filtering necessarily requires a physical gating of the signal before it is sent through the untrusted channel. 
Nonetheless, only investigating the potential advantage of time filtering, post-processing yields the same results as a physically gated signal stream.

To mimic this gating using post-selection, we consider only events in the auto-correlation measurement that occur within a post-selection window, $g^{(2)}(0)[\tau_{\text{A}}, t_0=0]$, where $\tau_{\text{A}}$ is the tuning parameter.
When calculating the resulting $g^{(2)}(0)$ value, we also apply the post-selection window to the $n$ uncorrelated peaks where $t_0=\pm\,nT_{\text{rep}}$ (see Supplementary~Note~\ref{sec:dataAna}).
Furthermore, we compute the corrected brightness as 

\begin{equation}
    B_{\text{corrected}}(\tau_{\text{A}}) = B\,\frac{A_{\text{uncorr}}(\tau_{\text{A}})}{A_{\text{uncorr}}(T_{\text{rep}})}
\end{equation}

where $A_{\text{uncorr}}$ denotes the average, blinking-corrected area of the uncorrelated peak.
Finally, the receiver could choose to disregard signals occurring outside an acceptance window~$\tau_{\text{B}}$ either by gating the single-photon detectors or by post-processing.
To satisfy the security requirements of the parameter estimation (see Supplementary~Note~\ref{sec:paramEst}), $\tau_{\text{B}}\geqslant\tau_{\text{A}}$ must hold.
For the following analysis, we set $\tau_{\text{B}}=\tau_{\text{A}}$ and assume the dark counts to be constant in time ($Y_0(t)=Y_0$) such that the time-filtered dark-count probability reads as
\begin{equation}
    Y_0(\tau_{\text{A}})=Y_0\frac{\tau_{\text{A}}}{T_{\text{rep}}}\,.
\end{equation}

\subsection{Time filtering for QKD}

\begin{figure*}
	\begin{center}
		\includegraphics[width=150mm]{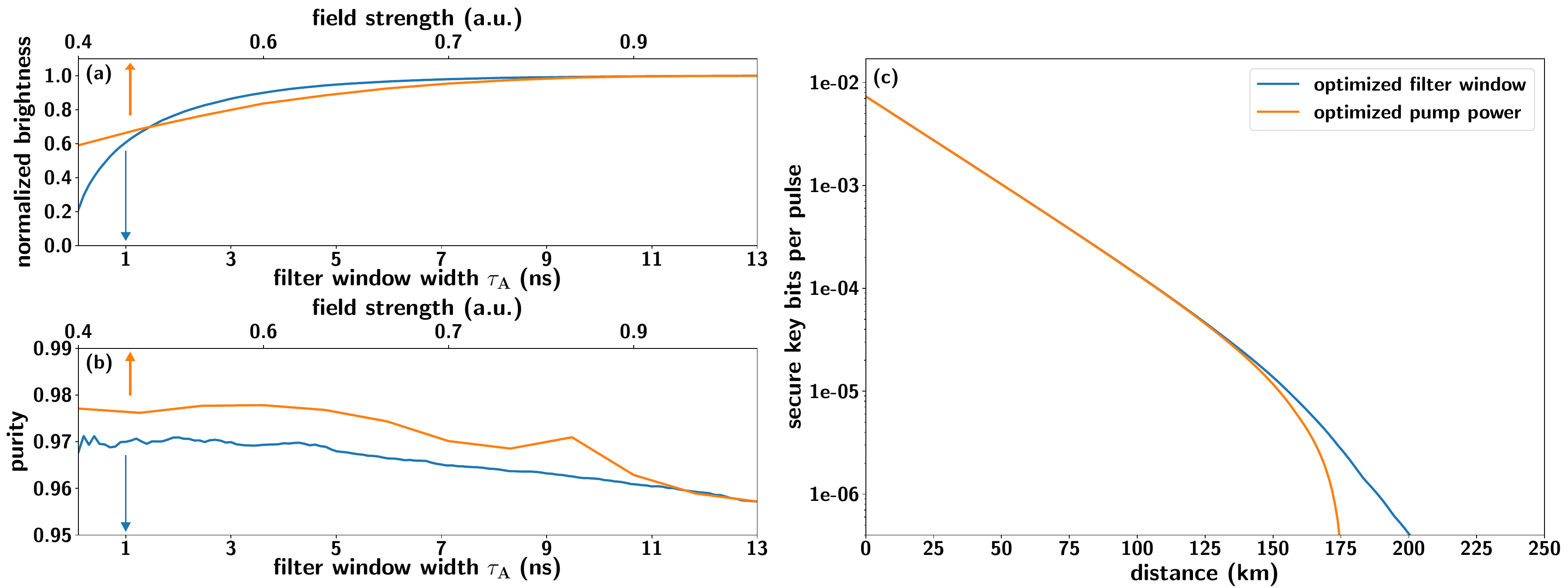}
		\caption{\textbf{Comparing time filtering and pump power tuning for QKD.} Both changing the excitation power or the width of a post-selection window $\tau_{\text{A}}$ alter the effective photon-number statistics of the QD source. For a laser detuning of $\Delta\lambda=1.5\,$nm, we analyze their effects in terms of brightness (a), single-photon purity (b) and secure key bits per pulse in a BB84 QKD protocol (c). When the power is chosen as tuning parameter, no time filtering is applied (i.e. $\tau_{\text{A}}=13.16\,$ns), whereas the field strength is set to $1\,$a.u. when varying $\tau_{\text{A}}$. The parameters for (c) are: single-photon detection error $e_d=0.02$, detection efficiency $\eta_d=0.86$, dark-count probability $Y_0=1.6\cdot10^{-6}$, error-correction code inefficiency $f=1.2$, fiber attenuation $\alpha=0.17\,$dB/km.}
		\label{fig:timefiltering}
	\end{center}
\end{figure*}

Supplementary~Note~\ref{fig:timefiltering} (a)-(b) shows the resulting brightness and purity as function of the window widths $\tau_{\text{A}}$ and -- for comparison -- as function of excitation power.
To asses the impact on a QKD implementation, we compute the $S\!K$ over distance and optimize for the $S\!K$ by tuning either $\tau_{\text{A}}$ or the pump power at each step (see Supplementary~Note~\ref{fig:timefiltering} (c)).
For simplicity, we restrict the analysis to a fixed detuning $\Delta\lambda=\SI{1.5}{\nano\meter}$.

While both methods perform similar for short to medium distances, we find an improvement of rate at large distances if time filtering is applied.
Interestingly, the enhanced $S\!K$ at high loss is attributed not to the improvement of purity but to a reduction of brightness and, even more important, reduction of dark-count probability.
For the optimal filter window at \SI{200}{\kilo\meter}, $\tau_{\text{A}}=\SI{0.2}{\nano\second}$, the purity is in fact lower than for power-tuning but the simultaneous decrease of dark counts to $Y_0(\tau_{\text{A}})=2.4\cdot10^{-8}$ -- almost two order of magnitudes lower than unfiltered -- preponderates.
As discussed in Supplementary~Note~\ref{sec:ExtendendAnalysis}, the $S\!K$ at long distances is ultimately given by the multi-photon population $p_{\text{m}}$ and dark-count probability $Y_0$ where $p_{\text{m}}$ is not just affected by the purity but also the brightness (see Eq.~\ref{eq:S_pm_bound}).

However, we remark that such a short filter window requires a modulator with $>10\,$GHz bandwidth and would introduce significant loss at all distances.
Alternative routes to decrease $Y_0$ include the technological advancement of SNSPDs in the long run and -- already feasible today -- the optimization of their biasing. Reducing the bias current (or voltage, depending on the model) of the SNSPDs affects the detection efficiency but will also lower the dark-count probability. This approach is especially appealing for long-distance communication and can be employed in combination with the power and detuning optimization of photon-number statistics in LA excitation. 

\end{appendices}

\clearpage

\begin{thebibliography}{10}
\expandafter\ifx\csname url\endcsname\relax
  \def\url#1{\texttt{#1}}\fi
\expandafter\ifx\csname urlprefix\endcsname\relax\def\urlprefix{URL }\fi
\providecommand{\bibinfo}[2]{#2}
\providecommand{\eprint}[2][]{\url{#2}}

\bibitem{MLL:NatPhot12}
\bibinfo{author}{Mart{\'{\i}}n-L{\'{o}}pez, E.} \emph{et~al.}
\newblock \bibinfo{title}{Experimental realization of shor{\textquotesingle}s
  quantum factoring algorithm using qubit recycling}.
\newblock \emph{\bibinfo{journal}{Nature Photonics}}
  \textbf{\bibinfo{volume}{6}}, \bibinfo{pages}{773--776}
  (\bibinfo{year}{2012}).
\newblock \urlprefix\url{https://doi.org/10.1038/nphoton.2012.259}.

\bibitem{Saggio2021}
\bibinfo{author}{Saggio, V.} \emph{et~al.}
\newblock \bibinfo{title}{Experimental quantum speed-up in
  reinforcement~learning agents}.
\newblock \emph{\bibinfo{journal}{Nature}} \textbf{\bibinfo{volume}{591}},
  \bibinfo{pages}{229--233} (\bibinfo{year}{2021}).
\newblock \urlprefix\url{https://doi.org/10.1038/s41586-021-03242-7}.

\bibitem{Centrone2021}
\bibinfo{author}{Centrone, F.}, \bibinfo{author}{Kumar, N.},
  \bibinfo{author}{Diamanti, E.} \& \bibinfo{author}{Kerenidis, I.}
\newblock \bibinfo{title}{Experimental demonstration of quantum advantage for
  {NP} verification with limited information}.
\newblock \emph{\bibinfo{journal}{Nat. Commun.}} \textbf{\bibinfo{volume}{12}}
  (\bibinfo{year}{2021}).
\newblock \urlprefix\url{https://doi.org/10.1038/s41467-021-21119-1}.

\bibitem{Pan:RevMod20}
\bibinfo{author}{Xu, F.}, \bibinfo{author}{Ma, X.}, \bibinfo{author}{Zhang,
  Q.}, \bibinfo{author}{Lo, H.-K.} \& \bibinfo{author}{Pan, J.-W.}
\newblock \bibinfo{title}{Secure quantum key distribution with realistic
  devices}.
\newblock \emph{\bibinfo{journal}{Rev. Mod. Phys.}}
  \textbf{\bibinfo{volume}{92}}, \bibinfo{pages}{025002}
  (\bibinfo{year}{2020}).
\newblock
  \urlprefix\url{https://link.aps.org/doi/10.1103/RevModPhys.92.025002}.

\bibitem{BAL:npjqi17}
\bibinfo{author}{Bedington, R.}, \bibinfo{author}{Arrazola, J.-M.} \&
  \bibinfo{author}{Ling, A.}
\newblock \bibinfo{title}{Progress in satellite quantum key distribution}.
\newblock \emph{\bibinfo{journal}{npj Quantum Inf.}}
  \textbf{\bibinfo{volume}{3}}, \bibinfo{pages}{30} (\bibinfo{year}{2017}).
\newblock \urlprefix\url{https://doi.org/10.1038/s41534-017-0031-5}.

\bibitem{BBR:PRL18}
\bibinfo{author}{Boaron, A.} \emph{et~al.}
\newblock \bibinfo{title}{Secure quantum key distribution over 421 km of
  optical fiber}.
\newblock \emph{\bibinfo{journal}{Phys. Rev. Lett.}}
  \textbf{\bibinfo{volume}{121}}, \bibinfo{pages}{190502}
  (\bibinfo{year}{2018}).
\newblock
  \urlprefix\url{https://link.aps.org/doi/10.1103/PhysRevLett.121.190502}.

\bibitem{NYC:NatComm23}
\bibinfo{author}{Neves, S.} \emph{et~al.}
\newblock \bibinfo{title}{Experimental cheat-sensitive quantum weak coin
  flipping}.
\newblock \emph{\bibinfo{journal}{Nat. Commun.}} \textbf{\bibinfo{volume}{14}},
  \bibinfo{pages}{1855} (\bibinfo{year}{2023}).
\newblock \urlprefix\url{https://doi.org/10.1038/s41467-023-37566-x}.

\bibitem{BBB:NC11}
\bibinfo{author}{Berl{\'\i}n, G.} \emph{et~al.}
\newblock \bibinfo{title}{Experimental loss-tolerant quantum coin flipping}.
\newblock \emph{\bibinfo{journal}{Nat. Commun.}} \textbf{\bibinfo{volume}{2}},
  \bibinfo{pages}{561} (\bibinfo{year}{2011}).
\newblock \urlprefix\url{https://doi.org/10.1038/ncomms1572}.

\bibitem{PJ+:natcomm14}
\bibinfo{author}{Pappa, A.} \emph{et~al.}
\newblock \bibinfo{title}{Experimental plug and play quantum coin flipping}.
\newblock \emph{\bibinfo{journal}{Nat. Commun.}} \textbf{\bibinfo{volume}{5}},
  \bibinfo{pages}{3717} (\bibinfo{year}{2014}).
\newblock \urlprefix\url{https://doi.org/10.1038/ncomms4717}.

\bibitem{SKS:NatComm23}
\bibinfo{author}{Schiansky, P.} \emph{et~al.}
\newblock \bibinfo{title}{Demonstration of quantum-digital payments}.
\newblock \emph{\bibinfo{journal}{Nat. Commun.}} \textbf{\bibinfo{volume}{14}}
  (\bibinfo{year}{2023}).
\newblock \urlprefix\url{https://doi.org/10.1038/s41467-023-39519-w}.

\bibitem{Kent:npjQI22}
\bibinfo{author}{Kent, A.}, \bibinfo{author}{Lowndes, D.},
  \bibinfo{author}{Pital{\'{u}}a-Garc{\'{\i}}a, D.} \& \bibinfo{author}{Rarity,
  J.}
\newblock \bibinfo{title}{Practical quantum tokens without quantum memories and
  experimental tests}.
\newblock \emph{\bibinfo{journal}{npj Quantum Inf.}}
  \textbf{\bibinfo{volume}{8}} (\bibinfo{year}{2022}).
\newblock \urlprefix\url{https://doi.org/10.1038\%2Fs41534-022-00524-4}.

\bibitem{GAA:pra18}
\bibinfo{author}{Guan, J.-Y.} \emph{et~al.}
\newblock \bibinfo{title}{Experimental preparation and verification of quantum
  money}.
\newblock \emph{\bibinfo{journal}{Phys. Rev. A}} \textbf{\bibinfo{volume}{97}},
  \bibinfo{pages}{032338} (\bibinfo{year}{2018}).
\newblock \eprint{1709.05882}.

\bibitem{NJC:NatComms12}
\bibinfo{author}{Ng, S.~K., N. Huei Y.and~Joshi}, \bibinfo{author}{Chen~Ming,
  C.}, \bibinfo{author}{Kurtsiefer, C.} \& \bibinfo{author}{Wehner, S.}
\newblock \bibinfo{title}{Experimental implementation of bit commitment in the
  noisy-storage model}.
\newblock \emph{\bibinfo{journal}{Nat. Commun.}} \textbf{\bibinfo{volume}{3}}
  (\bibinfo{year}{2012}).

\bibitem{Zbind:PRL13}
\bibinfo{author}{Lunghi, T.} \emph{et~al.}
\newblock \bibinfo{title}{Experimental bit commitment based on quantum
  communication and special relativity}.
\newblock \emph{\bibinfo{journal}{Phys. Rev. Lett.}}
  \textbf{\bibinfo{volume}{111}}, \bibinfo{pages}{180504}
  (\bibinfo{year}{2013}).
\newblock
  \urlprefix\url{https://link.aps.org/doi/10.1103/PhysRevLett.111.180504}.

\bibitem{Pan:PRL14}
\bibinfo{author}{Liu, Y.} \emph{et~al.}
\newblock \bibinfo{title}{Experimental unconditionally secure bit commitment}.
\newblock \emph{\bibinfo{journal}{Phys. Rev. Lett.}}
  \textbf{\bibinfo{volume}{112}}, \bibinfo{pages}{010504}
  (\bibinfo{year}{2014}).
\newblock
  \urlprefix\url{https://link.aps.org/doi/10.1103/PhysRevLett.112.010504}.

\bibitem{Kimble:2008}
\bibinfo{author}{Kimble, H.~J.}
\newblock \bibinfo{title}{The quantum internet}.
\newblock \emph{\bibinfo{journal}{Nature}} \textbf{\bibinfo{volume}{453}},
  \bibinfo{pages}{1023--1030} (\bibinfo{year}{2008}).
\newblock \urlprefix\url{https://doi.org/10.1038%2Fnature07127}.

\bibitem{Ren:2017}
\bibinfo{author}{Ren, J.-G.} \emph{et~al.}
\newblock \bibinfo{title}{Ground-to-satellite quantum teleportation}.
\newblock \emph{\bibinfo{journal}{Nature}} \textbf{\bibinfo{volume}{549}},
  \bibinfo{pages}{70--73} (\bibinfo{year}{2017}).
\newblock \urlprefix\url{https://doi.org/10.1038%2Fnature23675}.

\bibitem{Ma:2012}
\bibinfo{author}{Ma, X.-S.} \emph{et~al.}
\newblock \bibinfo{title}{Quantum teleportation over 143 kilometres using
  active feed-forward}.
\newblock \emph{\bibinfo{journal}{Nature}} \textbf{\bibinfo{volume}{489}},
  \bibinfo{pages}{269--273} (\bibinfo{year}{2012}).
\newblock \urlprefix\url{https://doi.org/10.1038/nature11472}.

\bibitem{Hensen:2015}
\bibinfo{author}{Hensen, B.} \emph{et~al.}
\newblock \bibinfo{title}{Loophole-free bell inequality violation using
  electron spins separated by 1.3 kilometres}.
\newblock \emph{\bibinfo{journal}{Nature}} \textbf{\bibinfo{volume}{526}},
  \bibinfo{pages}{682--686} (\bibinfo{year}{2015}).
\newblock \urlprefix\url{https://doi.org/10.1038/nature15759}.

\bibitem{Ritter:2012}
\bibinfo{author}{Ritter, S.} \emph{et~al.}
\newblock \bibinfo{title}{An elementary quantum network of single atoms in
  optical cavities}.
\newblock \emph{\bibinfo{journal}{Nature}} \textbf{\bibinfo{volume}{484}},
  \bibinfo{pages}{195--200} (\bibinfo{year}{2012}).
\newblock \urlprefix\url{https://doi.org/10.1038%2Fnature11023}.

\bibitem{Wang2019a}
\bibinfo{author}{Wang, H.} \emph{et~al.}
\newblock \bibinfo{title}{{Towards optimal single-photon sources from polarized
  microcavities}}.
\newblock \emph{\bibinfo{journal}{Nat. Photon.}} \textbf{\bibinfo{volume}{13}},
  \bibinfo{pages}{770--775} (\bibinfo{year}{2019}).
\newblock \urlprefix\url{http://www.nature.com/articles/s41566-019-0494-3}.

\bibitem{BassoBasset2019}
\bibinfo{author}{{Basso Basset}, F.} \emph{et~al.}
\newblock \bibinfo{title}{{Entanglement Swapping with Photons Generated on
  Demand by a Quantum Dot}}.
\newblock \emph{\bibinfo{journal}{Phys. Rev. Lett.}}
  \textbf{\bibinfo{volume}{123}}, \bibinfo{pages}{160501}
  (\bibinfo{year}{2019}).
\newblock \urlprefix\url{https://doi.org/10.1103/PhysRevLett.123.160501
  https://link.aps.org/doi/10.1103/PhysRevLett.123.160501}.

\bibitem{Zopf2019}
\bibinfo{author}{Zopf, M.} \emph{et~al.}
\newblock \bibinfo{title}{{Entanglement Swapping with Semiconductor-Generated
  Photons Violates Bell's Inequality}}.
\newblock \emph{\bibinfo{journal}{Phys. Rev. Lett.}}
  \textbf{\bibinfo{volume}{123}}, \bibinfo{pages}{160502}
  (\bibinfo{year}{2019}).
\newblock \urlprefix\url{https://doi.org/10.1103/PhysRevLett.123.160502
  https://link.aps.org/doi/10.1103/PhysRevLett.123.160502}.

\bibitem{Tomm:2021aa}
\bibinfo{author}{Tomm, N.} \emph{et~al.}
\newblock \bibinfo{title}{A bright and fast source of coherent single photons}.
\newblock \emph{\bibinfo{journal}{Nat. Nanotechnol.}}  (\bibinfo{year}{2021}).
\newblock \urlprefix\url{https://doi.org/10.1038/s41565-020-00831-x}.

\bibitem{Zhai2022}
\bibinfo{author}{Zhai, L.} \emph{et~al.}
\newblock \bibinfo{title}{{Quantum interference of identical photons from
  remote GaAs quantum dots}}.
\newblock \emph{\bibinfo{journal}{Nature Nanotechnology}}
  \textbf{\bibinfo{volume}{17}}, \bibinfo{pages}{829--833}
  (\bibinfo{year}{2022}).
\newblock \urlprefix\url{https://www.nature.com/articles/s41565-022-01131-2}.

\bibitem{AMH:npjQI20}
\bibinfo{author}{Anderson, M.} \emph{et~al.}
\newblock \bibinfo{title}{Quantum teleportation using highly coherent emission
  from telecom c-band quantum dots}.
\newblock \emph{\bibinfo{journal}{npj Quantum Inf.}}
  \textbf{\bibinfo{volume}{6}}, \bibinfo{pages}{14} (\bibinfo{year}{2020}).
\newblock \urlprefix\url{https://doi.org/10.1038/s41534-020-0249-5}.

\bibitem{BSS:npjQI21}
\bibinfo{author}{Basset, F.~B.} \emph{et~al.}
\newblock \bibinfo{title}{Quantum teleportation with imperfect quantum dots}.
\newblock \emph{\bibinfo{journal}{npj Quantum Inf.}}
  \textbf{\bibinfo{volume}{7}}, \bibinfo{pages}{7} (\bibinfo{year}{2021}).
\newblock \urlprefix\url{https://doi.org/10.1038/s41534-020-00356-0}.

\bibitem{Lodahl:2017}
\bibinfo{author}{Lodahl, P.}
\newblock \bibinfo{title}{Quantum-dot based photonic quantum networks}.
\newblock \emph{\bibinfo{journal}{Quantum Science and Technology}}
  \textbf{\bibinfo{volume}{3}}, \bibinfo{pages}{013001} (\bibinfo{year}{2017}).
\newblock \urlprefix\url{https://doi.org/10.1088/2058-9565/aa91bb}.

\bibitem{Liao2017}
\bibinfo{author}{Liao, S.~K.} \emph{et~al.}
\newblock \bibinfo{title}{{Long-distance free-space quantum key distribution in
  daylight towards inter-satellite communication}}.
\newblock \emph{\bibinfo{journal}{Nat. Photon.}} \textbf{\bibinfo{volume}{11}},
  \bibinfo{pages}{509--513} (\bibinfo{year}{2017}).
\newblock
  \urlprefix\url{http://www.nature.com/doifinder/10.1038/nphoton.2017.116}.

\bibitem{Wang2020}
\bibinfo{author}{Wang, J.}, \bibinfo{author}{Sciarrino, F.},
  \bibinfo{author}{Laing, A.} \& \bibinfo{author}{Thompson, M.~G.}
\newblock \bibinfo{title}{{Integrated photonic quantum technologies}}.
\newblock \emph{\bibinfo{journal}{Nat. Photon.}} \textbf{\bibinfo{volume}{14}},
  \bibinfo{pages}{273--284} (\bibinfo{year}{2020}).
\newblock \urlprefix\url{http://dx.doi.org/10.1038/s41566-019-0532-1
  http://www.nature.com/articles/s41566-019-0532-1}.

\bibitem{Miyazawa2016}
\bibinfo{author}{Miyazawa, T.} \emph{et~al.}
\newblock \bibinfo{title}{{Single-photon emission at 1.5 $\mu$ m from an
  InAs/InP quantum dot with highly suppressed multi-photon emission
  probabilities}}.
\newblock \emph{\bibinfo{journal}{Applied Physics Letters}}
  \textbf{\bibinfo{volume}{109}}, \bibinfo{pages}{132106}
  (\bibinfo{year}{2016}).
\newblock \urlprefix\url{http://dx.doi.org/10.1063/1.4961888
  http://aip.scitation.org/doi/10.1063/1.4961888}.

\bibitem{Anderson2020b}
\bibinfo{author}{Anderson, M.} \emph{et~al.}
\newblock \bibinfo{title}{{Gigahertz-Clocked Teleportation of Time-Bin Qubits
  with a Quantum Dot in the Telecommunication C Band}}.
\newblock \emph{\bibinfo{journal}{Physical Review Applied}}
  \textbf{\bibinfo{volume}{13}}, \bibinfo{pages}{054052}
  (\bibinfo{year}{2020}).
\newblock \urlprefix\url{https://doi.org/10.1103/PhysRevApplied.13.054052
  https://link.aps.org/doi/10.1103/PhysRevApplied.13.054052}.

\bibitem{Shooter2020}
\bibinfo{author}{Shooter, G.} \emph{et~al.}
\newblock \bibinfo{title}{{1GHz clocked distribution of electrically generated
  entangled photon pairs}}.
\newblock \emph{\bibinfo{journal}{Optics Express}}
  \textbf{\bibinfo{volume}{28}}, \bibinfo{pages}{36838} (\bibinfo{year}{2020}).
\newblock
  \urlprefix\url{https://opg.optica.org/abstract.cfm?URI=oe-28-24-36838}.

\bibitem{Lettner2021}
\bibinfo{author}{Lettner, T.} \emph{et~al.}
\newblock \bibinfo{title}{{Strain-Controlled Quantum Dot Fine Structure for
  Entangled Photon Generation at 1550 nm}}.
\newblock \emph{\bibinfo{journal}{Nano Letters}} \textbf{\bibinfo{volume}{21}},
  \bibinfo{pages}{10501--10506} (\bibinfo{year}{2021}).
\newblock
  \urlprefix\url{https://pubs.acs.org/doi/10.1021/acs.nanolett.1c04024}.

\bibitem{Sittig2022}
\bibinfo{author}{Sittig, R.} \emph{et~al.}
\newblock \bibinfo{title}{{Thin-film InGaAs metamorphic buffer for telecom
  C-band InAs quantum dots and optical resonators on GaAs platform}}.
\newblock \emph{\bibinfo{journal}{Nanophotonics}}
  \textbf{\bibinfo{volume}{11}}, \bibinfo{pages}{1109--1116}
  (\bibinfo{year}{2022}).
\newblock
  \urlprefix\url{https://www.degruyter.com/document/doi/10.1515/nanoph-2021-0552/html}.
\newblock \eprint{2107.13371}.

\bibitem{Nawrath2023}
\bibinfo{author}{Nawrath, C.} \emph{et~al.}
\newblock \bibinfo{title}{Bright source of purcell-enhanced, triggered, single
  photons in the telecom c-band}.
\newblock \emph{\bibinfo{journal}{Advanced Quantum Technologies}}
  \textbf{\bibinfo{volume}{n/a}}, \bibinfo{pages}{2300111}.
\newblock
  \urlprefix\url{https://onlinelibrary.wiley.com/doi/abs/10.1002/qute.202300111}.

\bibitem{vLeent2020}
\bibinfo{author}{van Leent, T.} \emph{et~al.}
\newblock \bibinfo{title}{Long-distance distribution of atom-photon
  entanglement at telecom wavelength}.
\newblock \emph{\bibinfo{journal}{Phys. Rev. Lett.}}
  \textbf{\bibinfo{volume}{124}}, \bibinfo{pages}{010510}
  (\bibinfo{year}{2020}).
\newblock
  \urlprefix\url{https://link.aps.org/doi/10.1103/PhysRevLett.124.010510}.

\bibitem{PhysRevLett.126.233601}
\bibinfo{author}{Thomas, S.~E.} \emph{et~al.}
\newblock \bibinfo{title}{Bright polarized single-photon source based on a
  linear dipole}.
\newblock \emph{\bibinfo{journal}{Phys. Rev. Lett.}}
  \textbf{\bibinfo{volume}{126}}, \bibinfo{pages}{233601}
  (\bibinfo{year}{2021}).
\newblock
  \urlprefix\url{https://link.aps.org/doi/10.1103/PhysRevLett.126.233601}.

\bibitem{Armando:ACS17}
\bibinfo{author}{Reindl, M.} \emph{et~al.}
\newblock \bibinfo{title}{Phonon-assisted two-photon interference from remote
  quantum emitters}.
\newblock \emph{\bibinfo{journal}{Nano Lett.}} \textbf{\bibinfo{volume}{17}},
  \bibinfo{pages}{4090--4095} (\bibinfo{year}{2017}).
\newblock \urlprefix\url{http://dx.doi.org/10.1021/acs.nanolett.7b00777}.

\bibitem{ReindlPRB2019}
\bibinfo{author}{Reindl, M.} \emph{et~al.}
\newblock \bibinfo{title}{Highly indistinguishable single photons from
  incoherently excited quantum dots}.
\newblock \emph{\bibinfo{journal}{Physical Review B}}
  \textbf{\bibinfo{volume}{100}} (\bibinfo{year}{2019}).
\newblock \urlprefix\url{https://doi.org/10.1103/physrevb.100.155420}.

\bibitem{BVC:npjQI22}
\bibinfo{author}{Bozzio, M.} \emph{et~al.}
\newblock \bibinfo{title}{Enhancing quantum cryptography with quantum dot
  single-photon sources}.
\newblock \emph{\bibinfo{journal}{npj Quantum Inf.}}
  \textbf{\bibinfo{volume}{8}}, \bibinfo{pages}{104} (\bibinfo{year}{2022}).
\newblock \urlprefix\url{https://doi.org/10.1038%2Fs41534-022-00626-z}.

\bibitem{Axt:PRL19}
\bibinfo{author}{Cosacchi, M.}, \bibinfo{author}{Ungar, F.},
  \bibinfo{author}{Cygorek, M.}, \bibinfo{author}{Vagov, A.} \&
  \bibinfo{author}{Axt, V.~M.}
\newblock \bibinfo{title}{Emission-frequency separated high quality
  single-photon sources enabled by phonons}.
\newblock \emph{\bibinfo{journal}{Phys. Rev. Lett.}}
  \textbf{\bibinfo{volume}{123}}, \bibinfo{pages}{017403}
  (\bibinfo{year}{2019}).
\newblock
  \urlprefix\url{https://link.aps.org/doi/10.1103/PhysRevLett.123.017403}.

\bibitem{Ma:Thesis08}
\bibinfo{author}{Ma, X.}
\newblock \bibinfo{title}{Quantum cryptography: theory and practice}
  (\bibinfo{year}{2008}).
\newblock \eprint{0808.1385}.

\bibitem{QBL:PRL2015}
\bibinfo{author}{Quilter, J.~H.} \emph{et~al.}
\newblock \bibinfo{title}{Phonon-assisted population inversion of a single
  $\mathrm{InGaAs}/\mathrm{GaAs}$ quantum dot by pulsed laser excitation}.
\newblock \emph{\bibinfo{journal}{Phys. Rev. Lett.}}
  \textbf{\bibinfo{volume}{114}}, \bibinfo{pages}{137401}
  (\bibinfo{year}{2015}).
\newblock
  \urlprefix\url{https://link.aps.org/doi/10.1103/PhysRevLett.114.137401}.

\bibitem{Bounouar2015}
\bibinfo{author}{Bounouar, S.} \emph{et~al.}
\newblock \bibinfo{title}{Phonon-assisted robust and deterministic two-photon
  biexciton preparation in a quantum dot}.
\newblock \emph{\bibinfo{journal}{Phys. Rev. B}} \textbf{\bibinfo{volume}{91}},
  \bibinfo{pages}{161302} (\bibinfo{year}{2015}).
\newblock \urlprefix\url{https://link.aps.org/doi/10.1103/PhysRevB.91.161302}.

\bibitem{Glaessl2013Pro}
\bibinfo{author}{Gl\"assl, M.}, \bibinfo{author}{Barth, A.~M.} \&
  \bibinfo{author}{Axt, V.~M.}
\newblock \bibinfo{title}{Proposed robust and high-fidelity preparation of
  excitons and biexcitons in semiconductor quantum dots making active use of
  phonons}.
\newblock \emph{\bibinfo{journal}{Phys.\ Rev.\ Lett.}}
  \textbf{\bibinfo{volume}{110}}, \bibinfo{pages}{147401}
  (\bibinfo{year}{2013}).
\newblock
  \urlprefix\url{https://journals.aps.org/prl/abstract/10.1103/PhysRevLett.110.147401 }.

\bibitem{Gustin.2019}
\bibinfo{author}{Gustin, C.} \& \bibinfo{author}{Hughes, S.}
\newblock \bibinfo{title}{Efficient pulse--excitation techniques for single
  photon sources from quantum dots in optical cavities}.
\newblock \emph{\bibinfo{journal}{Advanced Quantum Technologies}}
  \textbf{\bibinfo{volume}{3}}, \bibinfo{pages}{1900073}
  (\bibinfo{year}{2019}).
\newblock \urlprefix\url{https://doi.org/10.1002/qute.201900073 }.

\bibitem{Barth2016b}
\bibinfo{author}{Barth, A.~M.} \emph{et~al.}
\newblock \bibinfo{title}{Fast and selective phonon-assisted state preparation
  of a quantum dot by adiabatic undressing}.
\newblock \emph{\bibinfo{journal}{Phys. Rev. B}} \textbf{\bibinfo{volume}{94}},
  \bibinfo{pages}{045306} (\bibinfo{year}{2016}).
\newblock \urlprefix\url{https://link.aps.org/doi/10.1103/PhysRevB.94.045306 }.

\bibitem{GLLP04}
\bibinfo{author}{Gottesman, D.}, \bibinfo{author}{Lo, H.-K.},
  \bibinfo{author}{L\"{u}tkenhaus, N.} \& \bibinfo{author}{Preskill, J.}
\newblock \bibinfo{title}{Security of quantum key distribution with imperfect
  devices}.
\newblock \emph{\bibinfo{journal}{Quantum Info. Comput.}}
  \textbf{\bibinfo{volume}{4}}, \bibinfo{pages}{325--360}
  (\bibinfo{year}{2004}).

\bibitem{Hanschke.2018}
\bibinfo{author}{Hanschke, L.} \emph{et~al.}
\newblock \bibinfo{title}{Quantum dot single-photon sources with ultra-low
  multi-photon probability}.
\newblock \emph{\bibinfo{journal}{npj Quantum Inf.}}
  \textbf{\bibinfo{volume}{4}} (\bibinfo{year}{2018}).
\newblock \urlprefix\url{https://www.nature.com/articles/s41534-018-0092-0}.

\bibitem{BOV:npj18}
\bibinfo{author}{Bozzio, M.} \emph{et~al.}
\newblock \bibinfo{title}{Experimental investigation of practical unforgeable
  quantum money}.
\newblock \emph{\bibinfo{journal}{npj Quantum Inf.}}
  \textbf{\bibinfo{volume}{4}}, \bibinfo{pages}{5} (\bibinfo{year}{2018}).
\newblock \urlprefix\url{https://doi.org/10.1038/s41534-018-0058-2 }.

\bibitem{BB84}
\bibinfo{author}{Bennett, C.~H.} \& \bibinfo{author}{Brassard, G.}
\newblock \bibinfo{title}{Quantum cryptography: Public key distribution and
  coin tossing}.
\newblock In \emph{\bibinfo{booktitle}{Proc. IEEE International Conference on
  Computers, Systems and Signal Processing}}, vol.~\bibinfo{volume}{1},
  \bibinfo{pages}{175--179} (\bibinfo{address}{Bangalore, India},
  \bibinfo{year}{1984}).
\newblock
  \urlprefix\url{https://researcher.watson.ibm.com/researcher/files/us-bennetc/BB84highest.pdf}.

\bibitem{LMC:PRL05}
\bibinfo{author}{Lo, H.-K.}, \bibinfo{author}{Ma, X.} \& \bibinfo{author}{Chen,
  K.}
\newblock \bibinfo{title}{Decoy state quantum key distribution}.
\newblock \emph{\bibinfo{journal}{Phys. Rev. Lett.}}
  \textbf{\bibinfo{volume}{94}}, \bibinfo{pages}{230504}
  (\bibinfo{year}{2005}).
\newblock
  \urlprefix\url{https://link.aps.org/doi/10.1103/PhysRevLett.94.230504 }.

\bibitem{W:PRL05}
\bibinfo{author}{Wang, X.-B.}
\newblock \bibinfo{title}{Beating the photon-number-splitting attack in
  practical quantum cryptography}.
\newblock \emph{\bibinfo{journal}{Phys. Rev. Lett.}}
  \textbf{\bibinfo{volume}{94}}, \bibinfo{pages}{230503}
  (\bibinfo{year}{2005}).
\newblock
  \urlprefix\url{https://link.aps.org/doi/10.1103/PhysRevLett.94.230503}.

\bibitem{Waks:PRA02}
\bibinfo{author}{Waks, E.}, \bibinfo{author}{Santori, C.} \&
  \bibinfo{author}{Yamamoto, Y.}
\newblock \bibinfo{title}{Security aspects of quantum key distribution with
  sub-poisson light}.
\newblock \emph{\bibinfo{journal}{Phys. Rev. A}} \textbf{\bibinfo{volume}{66}},
  \bibinfo{pages}{042315} (\bibinfo{year}{2002}).
\newblock \urlprefix\url{https://link.aps.org/doi/10.1103/PhysRevA.66.042315}.

\bibitem{Gruenwald:19}
\bibinfo{author}{Grünwald, P.}
\newblock \bibinfo{title}{Effective second-order correlation function and
  single-photon detection}.
\newblock \emph{\bibinfo{journal}{New Journal of Physics}}
  \textbf{\bibinfo{volume}{21}}, \bibinfo{pages}{093003}
  (\bibinfo{year}{2019}).
\newblock \urlprefix\url{https://dx.doi.org/10.1088/1367-2630/ab3 ae0}.

\bibitem{Carmesin2018}
\bibinfo{author}{Carmesin, C.} \emph{et~al.}
\newblock \bibinfo{title}{{Structural and optical properties of
  InAs/(In)GaAs/GaAs quantum dots with single-photon emission in the telecom
  C-band up to 77 K}}.
\newblock \emph{\bibinfo{journal}{Physical Review B}}
  \textbf{\bibinfo{volume}{98}}, \bibinfo{pages}{125407}
  (\bibinfo{year}{2018}).
\newblock \urlprefix\url{https://link.aps.org/doi/10.1103/PhysRevB.98.125407}.

\bibitem{Ates2013}
\bibinfo{author}{Ates, S.} \emph{et~al.}
\newblock \bibinfo{title}{Improving the performance of bright quantum dot
  single photon sources using temporal filtering via amplitude modulation}.
\newblock \emph{\bibinfo{journal}{Scientific Reports}}
  \textbf{\bibinfo{volume}{3}} (\bibinfo{year}{2013}).
\newblock \urlprefix\url{https://doi.org/10.1038/srep01397}.

\bibitem{Kupko.2020}
\bibinfo{author}{Kupko, T.} \emph{et~al.}
\newblock \bibinfo{title}{Tools for the performance optimization of
  single-photon quantum key distribution}.
\newblock \emph{\bibinfo{journal}{npj Quantum Inf.}}
  \textbf{\bibinfo{volume}{6}}, \bibinfo{pages}{29} (\bibinfo{year}{2020}).
\newblock \urlprefix\url{https://www.nature.com/articles/s41534-020-0262-8}.

\bibitem{Paul2017}
\bibinfo{author}{Paul, M.} \emph{et~al.}
\newblock \bibinfo{title}{{Single-photon emission at 1.55 $\mu$ m from
  MOVPE-grown InAs quantum dots on InGaAs/GaAs metamorphic buffers}}.
\newblock \emph{\bibinfo{journal}{Applied Physics Letters}}
  \textbf{\bibinfo{volume}{111}}, \bibinfo{pages}{033102}
  (\bibinfo{year}{2017}).
\newblock \urlprefix\url{http://aip.scitation.org/doi/10.1063/1.4993935  }.

\bibitem{Dusanowski2022}
\bibinfo{author}{Dusanowski, {\L}.} \emph{et~al.}
\newblock \bibinfo{title}{{Optical charge injection and coherent control of a
  quantum-dot spin-qubit emitting at telecom wavelengths}}.
\newblock \emph{\bibinfo{journal}{Nat. Commun.}} \textbf{\bibinfo{volume}{13}},
  \bibinfo{pages}{748} (\bibinfo{year}{2022}).
\newblock \urlprefix\url{https://www.nature.com/articles/s41467-022-28328-2}.

\bibitem{LP:QIC07}
\bibinfo{author}{Lo, H.-K.} \& \bibinfo{author}{Preskill, J.}
\newblock \bibinfo{title}{Security of quantum key distribution using weak
  coherent states with nonrandom phases}.
\newblock \emph{\bibinfo{journal}{Quantum Info. Comput.}}
  \textbf{\bibinfo{volume}{7}}, \bibinfo{pages}{431–458}
  (\bibinfo{year}{2007}).
\newblock \urlprefix\url{https://dl.acm.org/doi/10.5555/2011832.2011838    }.

\bibitem{BLM:PRL00}
\bibinfo{author}{Brassard, G.}, \bibinfo{author}{L\"utkenhaus, N.},
  \bibinfo{author}{Mor, T.} \& \bibinfo{author}{Sanders, B.~C.}
\newblock \bibinfo{title}{Limitations on practical quantum cryptography}.
\newblock \emph{\bibinfo{journal}{Phys. Rev. Lett.}}
  \textbf{\bibinfo{volume}{85}}, \bibinfo{pages}{1330--1333}
  (\bibinfo{year}{2000}).
\newblock \urlprefix\url{https://link.aps.org/doi/10.1103/PhysRevLett.85.1330     }.

\bibitem{Morrison2023}
\bibinfo{author}{Morrison, C.~L.} \emph{et~al.}
\newblock \bibinfo{title}{Single-emitter quantum key distribution over 175 km
  of fibre with optimised finite key rates}.
\newblock \emph{\bibinfo{journal}{Nature Communications}}
  \textbf{\bibinfo{volume}{14}} (\bibinfo{year}{2023}).
\newblock \urlprefix\url{https://doi.org/10.1038/s41467-023-39219-5     }.

\bibitem{Lo2004}
\bibinfo{author}{Lo, H.-K.}, \bibinfo{author}{Chau, H.} \&
  \bibinfo{author}{Ardehali, M.}
\newblock \bibinfo{title}{Efficient quantum key distribution scheme and a proof
  of its unconditional security}.
\newblock \emph{\bibinfo{journal}{Journal of Cryptology}}
  \textbf{\bibinfo{volume}{18}}, \bibinfo{pages}{133--165}
  (\bibinfo{year}{2004}).
\newblock \urlprefix\url{https://doi.org/10.1007/s00145-004-0142-y    }.

\bibitem{Yin2020}
\bibinfo{author}{Yin, H.-L.} \emph{et~al.}
\newblock \bibinfo{title}{Tight security bounds for decoy-state quantum key
  distribution}.
\newblock \emph{\bibinfo{journal}{Scientific Reports}}
  \textbf{\bibinfo{volume}{10}} (\bibinfo{year}{2020}).
\newblock \urlprefix\url{https://doi.org/10.1038/s41598-020-71107-6  }.

\end{thebibliography}

\end{document}